\documentstyle[prl,twocolumn,aps,epsf,amsmath,amssymb,epsfig]{revtex}
\begin{document}

\twocolumn[
\hsize\textwidth\columnwidth\hsize\csname@twocolumnfalse\endcsname
\draft

\title{Metallicity and its low temperature behavior in dilute 2D 
carrier systems}
\author{S. Das Sarma and E. H. Hwang}
\address{Condensed Matter Theory Center, 
Department of Physics, University of Maryland, College Park,
Maryland  20742-4111 } 
\date{\today}
\maketitle

\begin{abstract}

We theoretically consider the temperature and density dependent
transport properties of semiconductor-based 2D carrier systems within
the RPA-Boltzmann transport theory, taking into account realistic
screened charged impurity scattering in the semiconductor. We derive a
leading behavior in the transport property, which is exact in the
strict 2D approximation and provides a zeroth order explanation for
the strength of metallicity in various 2D carrier systems.
By carefully comparing the calculated full nonlinear 
temperature dependence of 
electronic resistivity at low temperatures with the corresponding 
asymptotic analytic form obtained in the $T/T_F \rightarrow 0$ limit, 
both within the RPA screened charged impurity
scattering theory, 
we critically discuss the applicability of the linear temperature 
dependent correction to the low temperature
resistivity in 2D semiconductor structures. 
We find quite generally that for charged ionized impurity 
scattering screened by the electronic dielectric function 
(within RPA or its suitable generalizations including local field 
corrections), the resistivity obeys the asymptotic linear form only 
in the extreme low temperature limit 
of $T/T_F \le 0.05$.
We point out the experimental implications of our findings and discuss 
in the context of the screening theory the relative strengths of 
metallicity in different 2D systems.


\end{abstract}
\vspace{0.5cm}
]

\newpage

\section{introduction}

A great deal of attention has recently been focused on the
temperature dependence of carrier (both electrons and holes, depending
on whether the 2D system is n-doped or p-doped -- in this paper the
terminology ``electron'' or ``electronic'' generically refers to
electrons or holes depending on the 2D system being considered)
resistivity, $\rho(T)$,  
at low temperatures (and densities), following the pioneering
experimental report by Kravchenko and collaborators
\cite{Kravchenko94} that the measured low temperature $\rho(T)$ shows
very strong ``metallic'' temperature 
dependence (i.e. $\rho(T)$ increasing with $T$) at some intermediate
densities (the
so-called ``metallic'' or the 2D ``metal'' phase) eventually making a
transition to a strongly insulating state at low carrier densities
($n$).  
(At high electron densities $\rho(T)$ shows  weak temperature
dependence \cite{note1} similar to 3D metals.) 
In particular, $\rho(T)$ could increase by as much as a factor of three
in 2D Si MOSFETs (for $n \sim 10^{11}cm^{-2}$) as temperature changes
from 50 mK to a few K \cite{Kravchenko94,d2}. While this metallicity
(in this paper 
``metallicity'' or ``''metallic behavior'' will exclusively signify
the unusually strong $T$-dependence of $\rho(T)$ in the metallic phase
above the critical density at which the system makes a transition to
the manifestly insulating phase) is by far the strongest in $n$-Si
MOSFET 2D structures, the phenomenon has by now been observed (with
large quantitative variations in the strength of the metallicity) in
essentially all the existing low density 2D semiconductor systems 
\cite{d3,d4,d5,d6,d7,d8,d9,noh} such
as $p$-GaAs, $p$- and $n$-SiGe, Si on sapphire (SOS), 
$n$-GaAs, $n$-AlAs, etc. Our
work (we concentrate here on $n$-Si MOS, $p$-GaAs, and $n$-GaAs, as
representative 2D systems) presented in this paper deals with the
currently controversial issue of understanding this metallicity 
from a theoretical perspective. 
In particular, we use a conventional Fermi liquid theory approach in 
explaining the strong temperature dependence of $\rho(T)$ in the
metallic phase. 
We use the well-established RPA-Boltzmann transport
theory for calculating $\rho(T)$ for 2D carrier systems, taking into
account resistive scattering of the carriers by RPA-screened charged
impurity random potential.
The basic physical picture is that of a strongly temperature 
dependent effective disorder seen by the 2D electrons at low carrier 
densities due to the temperature dependent screening of charged 
impurity scattering which gives rise to the 
dominant resistive mechanism in semiconductors 
at low temperatures. We have earlier 
obtained qualitative agreement with experimental low
density $\rho(T)$ measured in Si-MOS \cite{DH1}, $p$-GaAs\cite{DH2},
SiGe\cite{senz}, Si-MOS with substrate  
bias\cite{lewalle}, and $n$-GaAs 2D structures\cite{lilly} using this
microscopic screening theory approach. 
For higher carrier densities, however, this screening theory is known
\cite{Ando_rmp} to 
provide an excellent quantitative description of 2D
carrier transport.

In this paper we consider, motivated by recent 
theoretical and experimental development, the leading-order 
temperature dependence of the 2D ``metallic'' resistivity in the 
low temperature, $T/T_F \rightarrow 0$, limit 
(where $T_F=E_F/k_B \propto n$ is the 2D 
Fermi temperature), and provide a qualitative explanation for the relative 
strength of metallicity in various 2D systems. Such a unifying 
qualitative explanation for the relative metallicity strengths 
in different materials has so far been lacking in the literature.
The 2D ``metallic'' phase is unusual in the sense that the usual 
3D metals do not exhibit strongly temperature dependent resistivity 
(unless of course there is a superconducting transition, not relevant 
for our consideration) at low temperatures ($T<5K$), 
the so-called Bloch-Gr\"{u}neisen regime where acoustic phonon
scattering (the main mechanism contributing to the temperature
dependence of resistivity in bulk metals) essentially freezes
out. Although phonon scattering plays a subtle (albeit secondary) role
\cite{DH2} 
in the metallic behavior of GaAs-based (both electron and hole) 2D
systems, theoretical calculations \cite{kawamura,DH2,lilly}
definitively show phonon scattering 
to be of little significance in the observed low temperature ($\le
1K$) 2D metallicity -- in fact, in Si MOS-based 2D electron systems,
where the 2D metallicity is most pronounced (and first observed),
phonon scattering plays no roles whatsoever in $\rho(T)$ for the
experimentally relevant regime of $T < 5K$. Phonon scattering effects,
which we have considered elsewhere \cite{kawamura,DH2,lilly} in
providing an explanation for the 
observed non-monotonicity in $\rho(T)$ at intermediate temperatures
($T \sim 1-5K$) in 2D $n$- and $p$-GaAs systems, are not included in
the current work since the focus of this paper is the behavior of
$\rho(T)$ as $T/T_F \rightarrow 0$ where phonons surely play no
roles. We consider only disorder scattering due to random charged
impurities (and surface roughness scattering, cf. see Sec. V)
in this work.

The strong temperature dependent ``metallic'' resistivity in low
density 2D systems arise, in our view, from an interplay in the
disorder scattering between finite temperature (or, even ``high''
temperature in the sense that $T/T_F \sim n^{-1}$ 
is not necessarily small
as it is in 3D metals and could actually be of order unity in low
density 2D systems for $T \sim 1K$) and density dependent 2D screening
properties as reflected in the dimensionless parameter 
$q_{TF}/2k_F \sim n^{-1/2}$
where $q_{TF}$ and $k_F$ are respectively the 2D Thomas-Fermi
screening wave vector and the Fermi wave vector \cite{Ando_rmp}. 
The fundamental
difference between the semiconductor-based 2D ``metals'' of interest
to us in this paper and the usual 3D metals is the great discrepancy
in the magnitudes of $T/T_F$ and $q_{TF}/2k_F$ in the two systems : In
3D metals $T/T_F \sim 10^{-4}$ for $T \sim 1K$ whereas $T/T_F \sim 0.1
-1$ in 2D semiconductor 
systems, and $q_{TF}/2k_F \approx 1 $ in 3D metals
whereas $q_{TF}/2k_F$ varies between 0.1 and 20 as carrier density is
changed in the 2D systems. In addition static screening has
qualitatively different wave vector dependence in 2D and 3D systems,
leading to the observed strong metallicity in various 2D systems.  
In 3D metals the low temperature resistivity arises almost entirely 
from temperature independent short range disorder scattering, which 
leads to exponentially suppressed, $O(e^{-T/T_F})$, temperature 
dependence in the resistivity, and any residual small temperature 
dependence in $\rho(T)$ is contributed by phonon scattering 
(which produces the well-known Bloch-Gr\"{u}neisen behavior, 
$\rho(T) \approx \rho_0 + A T^{5}$, where the temperature-independent 
contribution $\rho_0$ arises from short-range disorder scattering 
whereas the very weak temperature dependence characterized by the 
second term arises from highly suppressed phonon scattering at low 
temperatures). By contrast, low temperature transport (neglecting weak 
localization effects) in 2D metallic systems of interest to us is 
dominated mostly by screened disordered scattering 
[i.e., $\rho(T) = \rho_0 + \triangle \rho(T)$ with both $\rho_0$ and 
$\triangle \rho$ being determined essentially by disorder for 
$T \le 5K$], which can be strongly temperature dependent at low 
densities by virtue of large possible values of the relevant 
parameters $T/T_F$ ($\sim 1$) and $q_{TF}/2k_F$ ($\sim 10-20$) 
at low densities and temperatures in 2D semiconductor structures. 
All localization (as well as interaction effects beyond RPA)
effects are ignored \cite{note1} in this paper.

The paper is organized as follows. In section II we discuss a scaling
property (see Appendix A for the theory) of the Boltzmann theory
resistivity within the RPA screened charged impurity scattering model,
which provides a zeroth order qualitative explanation for the strength
of the transport metallicity in 2D carrier systems.
In section III we discuss the asymptotic low temperature ($T/T_F
\rightarrow 0$) behavior of the calculated resistivity in the
RPA-Boltzmann model comparing it quantitatively with the full
temperature dependent resistivity (in the same modal) in order to
estimate the regime of validity of the leading-order temperature
expansion of resistivity. In IV we consider the various solid state
physics effects (e.g. the {\it quasi}-2D nature of the semiconductor
layer, the long-range or the short-range nature of the {\it bare}
scattering potential) on the 2D transport properties. In section V we
provide a critical comparison between our realistic (but theoretically
approximate) RPA-Boltzmann 2D transport theory and a set of recent
experimental results in Si inversion layer, concluding that our theory,
without any adjustment of parameters and/or {\it ad hoc} theoretical
refinement, describes well the observed experimental temperature
dependence down to a carrier density of about $5 \times 10^{11}
cm^{-2}$ --- for lower densities the agreement between experiment and
theory is at best qualitative with the actual temperature dependence
of $\rho(T)$ being stronger than the calculated $\rho(T)$. 
We conclude in VI with a discussion of
the implications of our results.

\section{Density-Temperature ($\lowercase{q}_0$, $\lowercase{t}$)
  scaling of metallicity}

In the RPA-Boltzmann theory (cf. Appendix A)
the dimensionless $r_s$ parameter 
(the so-called Wigner-Seitz radius)
characterizing the electron-electron interaction strength in the 
2D system does {\it not} play a fundamental role
in determining the temperature 
dependence of $\rho(T)$ except so far as $r_s$ determines the 
dimensionless parameters $T/T_F$ ($\equiv t$) and $q_{TF}/2k_F$
($\equiv q_0$) 
through the carrier density. 
We believe that the fundamental minimal parameters determining the
zeroth order 2D
metallicity, i.e. the temperature dependence of $\rho(T,n)$ in the
putative metallic phase, are $t$ and $q_0$. In particular, 
for 2D systems it is easy to show that $q_0 = g_{v}^{3/2}r_s /\sqrt{2}$ 
and $t = (k_B T /Ry) (g_{v}/2) r_s^2$, where $g_{v}$ is the valley
degeneracy of the relevant semiconductor material ($g_{v}=2$ for
Si(100)-MOS structures, $g_{v}=1$ for $p$- and $n$-GaAs systems),
$r_s = (\pi n)^{-1/2}/a_B$ is the usual dimensionless Wigner-Seitz
density (or interaction) parameter with $a_B=\kappa \hbar^2/me^2$ as
the effective semiconductor Bohr radius ($\kappa$ and $m$ are the
background dielectric constant and the carrier effective mass
respectively), and $Ry = e^2/(2\kappa a_B)$ is the natural atomic
energy unit (effective Rydberg) for the semiconductor. We find that
the existing experimental data for the metallicity in various 2D
semiconductor systems approximately obeys the 2-parameter scaling
behavior $\triangle \rho(T,n)/\rho_0 \sim F(t,q_0)$, where $F$ is a
smooth and approximately universal function of $q_0$ and $t$ for all
2D ``metallic'' systems, as implied by the
screening theory. In particular, a direct consequence of this
theoretical prediction is that the temperature dependence of $\rho(T)$
should correlate with the parameter $q_0$ 
in different materials when expressed in
terms of the dimensionless temperature $t$. (We note that the
functional dependence of  
$q_0 \sim g_{v}^{3/2} r_s$ is not only different from the
dimensionless density parameter $r_s$, but also from the dimensionless
ratio of the Fermi energy to the Coulomb energy in the system which
goes as $g_{v} r_s$.) 
Theoretical details and the equations for our 2D RPA-Boltzmann theory are
given in the Appendix A of this paper where we derive this scaling law.
We note that the scaling behavior $\triangle \rho/\rho_0 \sim
F(t,q_0)$, with $t = T/T_F$ and $q_0 = q_{TF}/2k_F$, derived in this
paper is {\it exact} for RPA screened charged impurity scattering in
the ideal 2D limit in contrast to various other scaling behaviors
(e.g.  refs. \onlinecite{DH1,DH2,lewalle}) discussed earlier in the
literature which are all approximate scaling behavior valid only in
limited range of parameters.

This zeroth order functional dependence on $t$ and $q_0$, $\triangle
\rho(T,n) \equiv \triangle \rho(t,q_0)$, in fact, provides a minimal
explanation for the observed strong variation in the 2D metallicity
not only for various densities in the same material but for different
materials at equivalent densities -- for example, Si-MOSFET based 2D
electron systems manifest much stronger metallicity compared with
$p$-GaAs or $n$-GaAs based 2D systems even at the same $r_s$  
value because $g_{v}=2$ (1) in Si (GaAs). 
Taking $r_s$ to be the critical parameter determining metallicity
fails to explain why the 2D $p$-GaAs system shows weaker metallic
behavior, even though it typically has much
larger $r_s$ values ($r_s \sim 15-40$) than the Si system. The
screening theory provides a simple explanation for this observation by
virtue of $q_0$ being larger (by a large factor of $\sqrt{8}$) in Si
than in GaAs for the same $r_s$ value. Thus, Si MOSFET 2D
metallicity at $r_s=10$ should be approximately comparable to a GaAs
2D metallicity at $r_s = 10 \sqrt{8} \approx 28$ when expressed as a
function of the dimensionless temperature variable $t=T/T_F$. This
predicted correspondence in the relative metallicity in terms of $q_0$
and $t$ is consistent with the experimental observations in these
systems. We emphasize, however, that this correspondence is expected
to work only on a zeroth-order qualitative level and should not, by
any means, be construed as a precise quantitative prediction. 
This is discussed with theoretical details in the appendix.
For example, the form factor effects associated with the quasi-2D
subband quantization do not scale with the density parameter $q_0$,
and will necessarily affect different systems in different manners
since the associated effective masses, the dielectric constants, the
depletion charge densities, and the confinement potentials, which
together determine the form-factor, are different in different
systems. Similarly, the bare (i.e., unscreened) disorder will
certainly depend on the system varying qualitatively among different
systems and materials, which could lead to substantial quantitative
deviations from our predicted $\rho(q_0,t)$ scaling based on the
temperature dependent screening argument. For example, in Si MOSFETs
transport is dominated by interface scattering -- both by long-range
potential scattering due to ionized impurities located near the  
Si-SiO$_2$ interface and short-range scattering by interface roughness
fluctuations inherent at the Si-SiO$_2$ interface. In GaAs structures
(both $p$ and $n$), scattering by (unintentional) background charge
impurities (and less so by remote charge dopants) dominates at the
highest mobilities with phonon scattering being non-negligible
(although small) down to $T <1K$. 
In addition, $p$-GaAs structures most likely also have significant
spin inter-subband scattering within the spin-split valence bands. The
non-universal  
quantitative effects associated with subband quantization and disorder
potential make the simple 2-parameter screening picture of $\triangle
\rho(T,n)$ dependent only on $q_0$ and $t$ quantitatively inaccurate,
but the simple picture applies surprisingly well on a zeroth-oder
qualitative level as can be verified by comparing the experimentally
observed metallicity strengths in Si MOS, $p$-GaAs, and $n$-GaAs
structures where the metallicity  scales approximately with
$q_0$ and $t$ provided $T$ is low enough so
that phonon effects could be ignored in the GaAs system.  
The crucial new point we are making in this paper is that strong
metallicity manifests itself in low density 2D systems because the
control parameters $q_0 \propto n^{-1/2}$ and $t \propto n^{-1}$ are
large only for low carrier densities and not because $r_s$ is large --
for example, n-Si MOS system and n-GaAs system show more than an order
of magnitude different metallicities for the same $r_s$ value because
$q_0$ and $t$ are much larger in Si than in n-GaAs ($\Delta
\rho/\rho_0$ increases by a factor of 3 in Si MOS systems of
refs. \onlinecite{Kravchenko94} and \onlinecite{lewalle} for $r_s \ge
10$ whereas it increases only by about 25\% in the n-GaAs system of
refs. \onlinecite{lilly} for $r_s \ge 10$).
Within a specific materials system, however, the $q_0$ dependence of
the resistivity becomes completely equivalent to an $r_s$ dependence
(since $g_{\nu}$ is a constant for a given system, and $q_0 \equiv
g_{\nu}^{3/2}r_s/\sqrt{2}$) as one would expect --- it is only in


\begin{figure}
\epsfysize=2.7in
\centerline{\epsffile{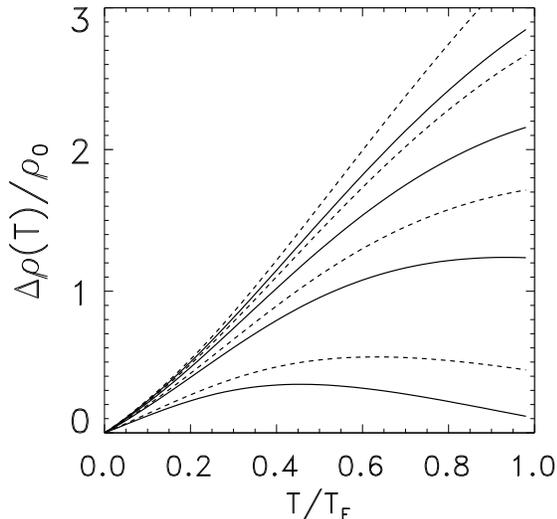}}
\caption{Calculated $\Delta \rho = \rho(T,n)-\rho_0$,
  where $\rho_0=\rho(T=0)$, for various
$q_0=q_{\rm TF}/2k_F=$ 5, 10, 15, 20 (from the bottom) 
as a function of $t=T/T_F$ for Si MOSFET.
The solid (dashed) lines indicate resistivities from interface (bulk)
charged impurities. }
\label{fig1}
\end{figure}
\noindent
comparing different systems (e.g. Si and GaAs) that $q_0$ and $r_s$
dependence are not equivalent.

In Fig. \ref{fig1} we show our calculated $\rho(T,n)$, depicted in
terms of the dimensionless variables $q_0$ and $t$, for a strictly 2D
system using Si MOS 
parameters (e.g., effective mass, dielectric constant, etc.) and
assuming the disorder scattering to be entirely due to finite
temperature RPA-screened charged impurity scattering.
Results shown in Fig. \ref{fig1} demonstrate the importance of the
dimensionless screening parameter $q_0$ ($=q_{\rm TF}/2k_F$) as the
relevant control parameter in determining the strength of metallicity
in 2D systems. In particular the maximum relative change in the
resistivity, $\triangle \rho/\rho_0$, ranges from about 40\% for
$q_0=5$ (which corresponds to a Si (100) inversion layer carrier
density of $n=11\times 10^{11}$ cm$^{-2}$, $T_F=80K$) to almost 300 \% for
$q_0=20$ (corresponding to $n=7\times 10^{10}$ cm$^{-2}$, $T_F = 5K$) 
as $t\equiv
T/T_F$ changes from 0 to 1 in Fig. \ref{fig1}. 
Therefore, the RPA results of Fig. 1 indicate an
increase of resistivity by about 10\% and 300\% respectively for
carrier densities $1.1\times 10^{12}cm^{-2}$ and $7\times
10^{10}cm^{-2}$ in Si MOSFETs for a change in $T$ of $0-5K$, assuming
that the system remains metallic. It should be noted that the
maximum in $\triangle \rho/\rho_0$ shifts to higher values of $t$
($=T/T_F$) for higher (lower) values of $q_{\rm TF}/2k_F$ ($n$), and
phonon effects (ignored in our consideration) will play increasingly
important quantitative role in the temperature dependent resistivity
for $T>5K$. These two facts together make the metallic behavior
relatively even more important at lower densities, or equivalently,
higher values of $q_0$.

A comparison between solid and dashed lines in
Fig. \ref{fig1} shows the quantitative importance of the nature of impurity
scattering in determining the temperature dependence of resistivity:
in general, charged impurities randomly distributed in the 2D layer
itself (bulk disorder shown by dashed lines in Fig. \ref{fig1}) lead to
stronger temperature dependence than interface disorder (solid lines)
associated with charged impurities distributed randomly at the
semiconductor-insulator interface. This is precisely what is expected
since screening effects should be the strongest when charged
impurities and the carriers are distributed in the same region of
space with no spatial separation. It may also be worthwhile to mention
in the context of Fig. \ref{fig1} that the experimental Si inversion layer
systems \cite{Kravchenko94,d2} manifesting the most dramatic
metallicity ({\it i.e.} large changes in $\triangle \rho/\rho_0$ as a
function of temperature) typically have $n\le 10^{11} cm^{-2}$
corresponding to $q_0 \sim 15-20$ in Fig. \ref{fig1}, thereby showing
a relative temperature dependent change in resistivity of about 100
--- 300 \% as $T/T_F$ varies from zero to 0.5. Thus the results in
Fig. \ref{fig1} are in reasonable 
qualitative agreement with experimental results
as we have emphasized elsewhere \cite{DH1}. 
Scattering mechanism not included in the theory (e.g. surface
roughness scattering) and higher-order interaction effects will
certainly modify the quantitative details of the results, but it is
gratifying to see that a zeroth-order Boltzmann transport theory
including only RPA screened charged impurity scattering provides a
reasonable qualitative description of the observed metallicity.

\begin{figure}
\epsfysize=3.in
\centerline{\epsffile{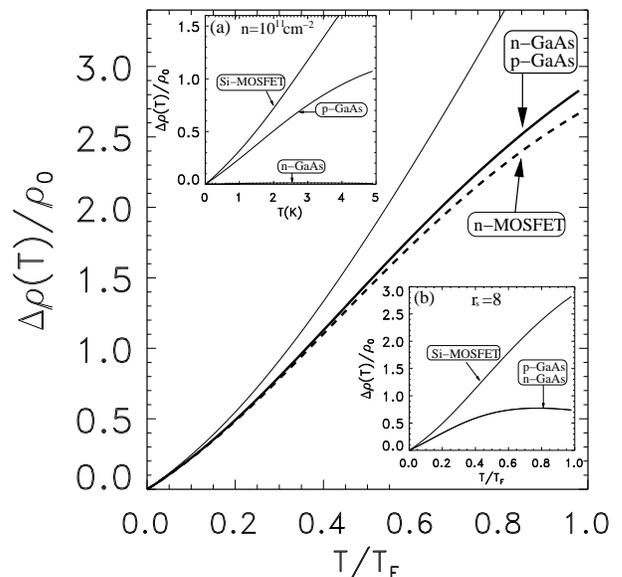}}

\caption{The main figure shows the scaled resistivity as a function of 
$t=T/T_F$ for Si-MOSFET, p-GaAs, and n-GaAs with a fixed $q_0=15$.
For pure 2D system the calculated resistivities yield perfect scaling
for all systems (thin solid line). Thick lines show the scaling for
quasi-2D systems.
In inset (a) the comparison of the
metallicity for fixed density $n=10^{11}cm^{-2}$ is given as a
function of a temperature. 
In inset (b) we show the change of resistivity for fixed value of
$r_s=8$ as a function of $t=T/T_F$. 
}
\label{fig2}
\end{figure}

In Fig. \ref{fig2}  we show the calculated $\triangle \rho/\rho_0$ for three
different 2D systems for a comparison of their metallicity: 
(100) n-Si inversion layer, p-GaAs heterostructure, and 

\noindent
n-GaAs
heterostructure. We have shown the results for the same values of
carrier density, $r_s$ parameter, and $q_0$ ($=q_{\rm TF}/2k_F$)
parameter in order to emphasize the importance of the dimensionless
screening parameter $q_0$ in determining the temperature
dependence. As emphasized above, for the same values of carrier
density (see inset (a)), the metallicity is the strongest (weakest)
in the Si (n-GaAs) system with the p-GaAs system being
intermediate. 
For the same value of $r_s$ parameter (see inset (b)) 
the metallicity depends strongly
on the valley degeneracy, but weakly on the other parameters. 
The mass of the carrier dose not change the  metallicity for the same
$r_s$ parameter, which gives exactly the same resistivity behavior
for p-GaAs and n-GaAs
if we use the same material parameters except mass. 
In the main figure we show the metallicity for equivalent $q_0=15$
parameter values in all three systems.
The thin solid line shows that the temperature dependence is
scaled perfectly for the pure 2D systems, as proven in the appendix.
For quasi-2D systems, however, the scaled resistivity 
expressed as a
function of dimensionless temperature $t=T/T_F$ is
approximately similar in the three systems without perfect scaling. 
The deviation of 
scaling is mostly due to form factor effects.
Since the electron
effective mass in n-GaAs is small ($m=0.067m_e$), extremely low values
of carrier density are required in 2D n-GaAs systems for observing
appreciable screening-induced temperature dependence as has been
recently reported in Ref. \onlinecite{lilly}. 


\section{Low temperature asymptotic behavior}

Since the behavior of $\triangle \rho(T,n)$ for finite $T$ is
necessarily non-universal for reasons discussed above, recent attention
has focused on the very low temperature behavior of $\rho(T)$ in the
$t=T/T_F \rightarrow 0$ limit. In particular, it was realized a long
time ago \cite{stern,gold}
that $\triangle \rho(T,n)$ derived from the Boltzmann theory
within the relaxation time approximation for RPA screened disorder
scattering has the following expansion in 2D systems: 
\begin{equation}
\triangle \rho(T,n)/\rho_0 \approx C_1 t + C_{3/2} t^{3/2} + \cdot
\cdot \cdot,
\end{equation}
in the asymptotic $t \rightarrow 0$ regime. The coefficients $C_1$
and $C_{3/2}$ were first (but incorrectly) calculated by Gold and
Dolgopolov \cite{gold}. We have re-calculated $C_1$ and $C_{3/2}$, and
find the 
original results in Ref.\onlinecite{gold} to be incorrect. We
calculate the correct coefficients (for RPA screening) to be: 
\begin{subequations}
\begin{eqnarray}
C_1 & = & 2 \left ( 1 + 1/q_0 f \right )^{-1}, \\
C_{3/2} & = & 2.646 \left (1+1/q_0 f \right )^{-2},
\end{eqnarray}
\end{subequations}
where $q_0 = q_{TF}/2k_F$ (as defined above) and $f \equiv f(2k_F)$ is
the appropriate quasi-2D subband form-factor at the wave vector $2k_F$
(in general $f\le 1$ with the strictly 2D limit being $f=1$, see
appendix A). 
Our calculated $C_1$ agrees (in the strictly 2D limit of $f=1$) with
the recent (Hartree) result given in Ref. \onlinecite{zala} and
disagree with
that of Ref. \onlinecite{gold} whose incorrectly calculated $C_1$ is
larger by a factor of $2\ln2$ (i.e. about $40\% $ higher).  
Our calculated $C_{3/2} \approx 2.65$ is $28 \%$ smaller in magnitude
than the incorrect value ($\approx 3.4$) quoted in
Ref. \onlinecite{gold}. We note that the errors in
Ref.\onlinecite{gold} led to a large overestimate of the asymptotic 
temperature
dependence of $\rho(T)$ in the Gold-Dolgopolov formula\cite{gold},
which we now correct. An important recent 
theoretical development in the subject has been the demonstration by
Zala {\it et al.} \cite{zala} that the leading-order linear result given in
Eq. 1, in fact, survives (albeit with $C_1$ replaced by an unknown
Fermi liquid parameter) inclusion of higher-order electron-electron
interaction terms in the theory, of which screening is only one
particular aspect. In particular, the leading-order temperature
dependence in the interaction theory of Zala {\it et al.} contains the $C_1$
term of our Eqs. (1) and (2) as the so-called Hartree term in the
language of Ref. \cite{zala}.

A thorough understanding of this first order linear term in the theory
has taken on significance in view of the existence of the
Zala {\it et al.} \cite{zala} work, and even more importantly, because
of the several recent attempts \cite{noh,pudalov,noh1}
to compare experimental results to the interaction theory.
The interaction theory
considerably extends the screening theory (through the inclusion of
higher-order interaction corrections), but is unfortunately
constrained at this stage to only the leading order result in $t$, and
therefore applies only at very high (low) densities (temperature) so
that the constraints $t \ll 1$ and $\triangle \rho/\rho_0 \ll 1$ are
satisfied.   
The two theories are thus complementary -- the screening theory being 
an approximate theory (because it includes only the screening part of
the electron-electron interaction) for {\it all} $t$ (in fact, its
accuracy improves with increasing $t$ since the RPA becomes exact in
the classical high-temperature limit) whereas the interaction theory
is presumably an exact leading-order in $t$ 
(within the perturbative Landau Fermi liquid theory scheme) theory as
$t\rightarrow 0$. This obvious complementarity of these two approaches
has not been emphasized in the recent literature where some recent
publications have even presented the misleading and incorrect viewpoint of 
these two approaches as mutually exclusive competing theories.
It is important to emphasize that the interaction theory \cite{zala},
by construction, applies only when the temperature correction to the
$T=0$ conductivity is small, i.e. the theory of ref. \onlinecite{zala}
is a leading order theory for small temperature corrections to the
$T=0$ conductivity as $T/T_F \rightarrow 0$. By definition, therefore,
this interaction theory cannot explain the strong metallicity or the
large temperature-dependent changes in the resistivity reported in the
literature. 
The real significance of ref. \onlinecite{zala} is strictly
theoretical --- it shows that the ``metallic'' behavior given by
Eq. (1) within the RPA-Boltzmann theory survives higher-order
electron-electron interaction corrections in the $T/T_F \rightarrow 0$
limit (provided that the ``ballistic'' transport condition $k_BT \gg
\hbar/\tau$, where $\tau$ is the $T=0$ transport relaxation time
($\tau = m/ne^2\rho_0$), is
satisfied, i.e., the temperature is in the intermediate range $T_F \gg
T \gg \hbar/\tau k_B$). We note that the necessary condition minimally
required for a comparison between experiment and the interaction
theory is that (1) the experimental conductivity must show a linear
temperature correction in the intermediate temperature regime $T_F \gg
T \gg \hbar/\tau k_B$, and (2) the temperature correction to the $T=0$
conductivity must be small. Most experiments on 2D transport do not
satisfy these necessary conditions for comparison with the interaction
theory, most particularly because the measured conductivity essentially
never manifests a linear temperature dependence except at very high
densities where the RPA-Boltzmann semiclassical transport theory is
quantitatively accurate.

In view of the complementary nature of these two theories 
it becomes important to ask 
about the regime of validity of the linear approximation inherent in
the interaction theory. This issue becomes particularly crucial since
most of the existing $\rho(T)$ data in the putative 2D metallic phase does
{\it not} follow a linear temperature dependence over any appreciable
temperature regime in the lowest temperature range (i.e. $T/T_F \ll
1$). The situation becomes more complicated with the realization that
the `strong' condition (the `weak' condition being $T\ll T_F$) for the
validity of the interaction (as well as the screening) theory is that
$T_D \ll T \ll T_F$, where $T_D \approx \hbar/(\tau k_B)$ is roughly
the so-called Dingle temperature with $\tau$ being the $T=0$ limit of
the transport relaxation time [{\it i.e.} $\tau=m/(ne^2\rho_0)]$. 
In the screening theory $T_D$ cuts off
the temperature dependence of screening for $T\le T_D$ (making the
disorder to be effectively temperature independent for $T\ll T_D$)
leading to a 
suppression in the temperature dependence of $\rho(T)$ whereas
the interaction theory is by construction a theory of ``ballistic''
transport developed in the $\hbar/(\tau k_B T) \ll 1$ limit (and then
the $T/T_F \ll 1$ limit is taken to obtain explicit asymptotic
results). The $T_D$-cutoff in the temperature dependence of $\rho(T)$
for $T\le T_D$ is extremely well-motivated physically within the
screening theory, and has been discussed in details in the literature
\cite{DH1,DH2,stern}
as the reason for the need of low disorder (or equivalently, 
high mobility with concomitant low values of $T_D$) samples to observe
2D metallicity (low density is also required in the screening theory
for strong metallicity so as to produce large values of $T/T_F$ at low
temperatures and to make $q_{TF}/2k_F$ large enough to have strong screening
effect). We will mostly ignore the Dingle temperature effects in the
theory by assuming $T_D \approx 0$, but in comparing experimental data
to the interaction theory the ballistic limit is an important
constrain to remember. 
In particular, in many experimental situations the constraint
$\hbar/(\tau k_B) \ll T \ll T_F$ necessary for the application of the
ballistic limit inter-


\begin{figure}
\epsfysize=2.7in
\centerline{\epsffile{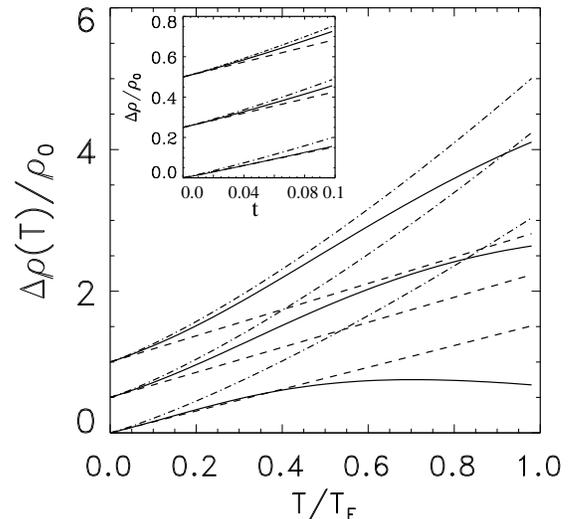}}
\caption{
$\rho(T)$ for a Si-MOSFET with 
density $n=$56.0, 12.0, 7.0$\times 10^{10}cm^{-2}$ (from the bottom).
Solid lines indicate the full RPA screening
theory, dashed lines the asymptotic approximations with only 
the linear term, and dot-dashed lines with both the $O(T)$ and
$O(T^{3/2})$ terms.  
For clarity we use offset by 0.5 for lower densities.
The inset shows $\rho(T)$ on an expanded temperature scale. 
}
\label{fig3}
\end{figure}
\noindent
action theory may not even exist \cite{noh1}.

\begin{figure}
\epsfysize=3.in
\centerline{\epsffile{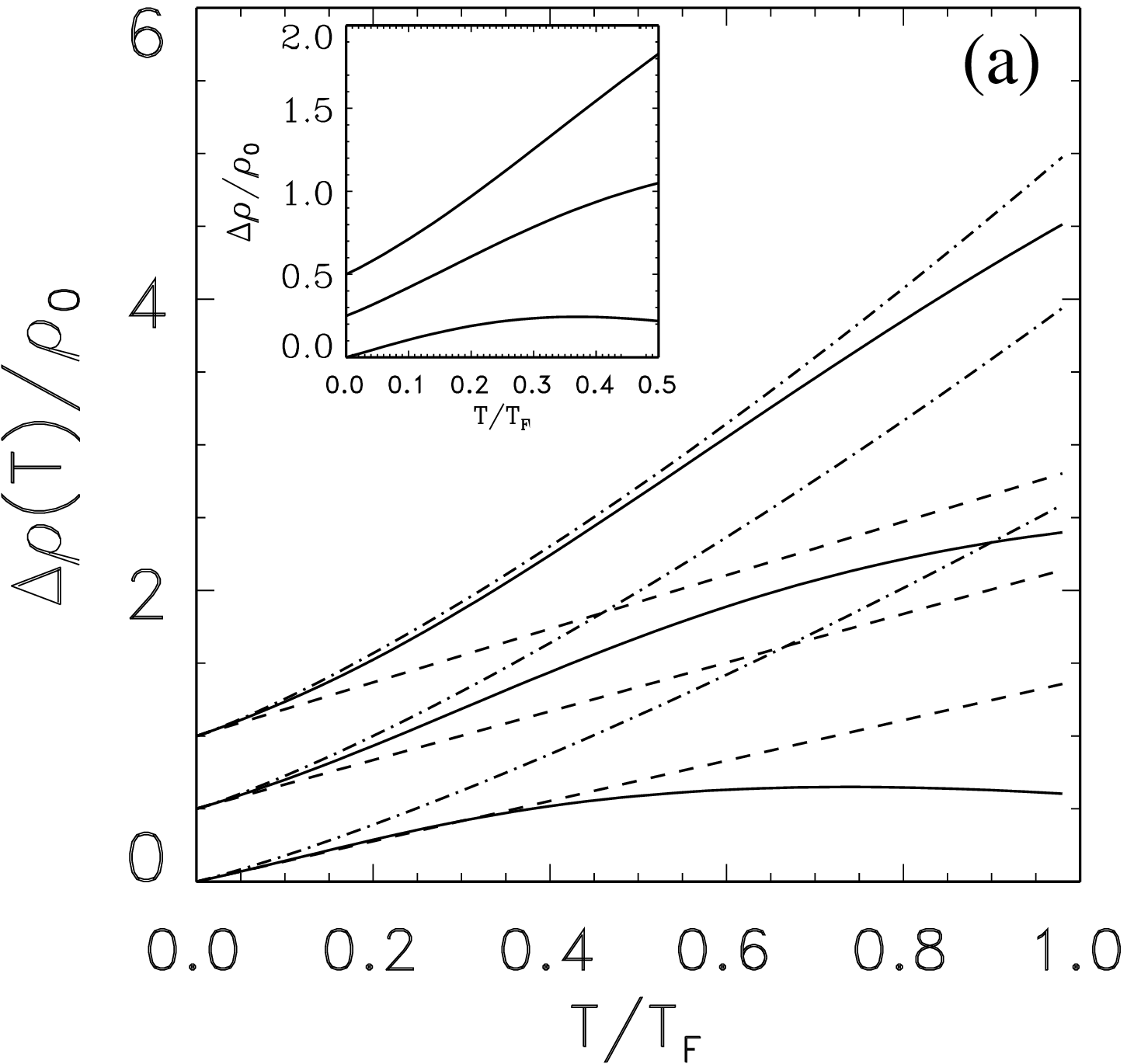}}
\vspace{0.5cm}
\epsfysize=3.in
\centerline{\epsffile{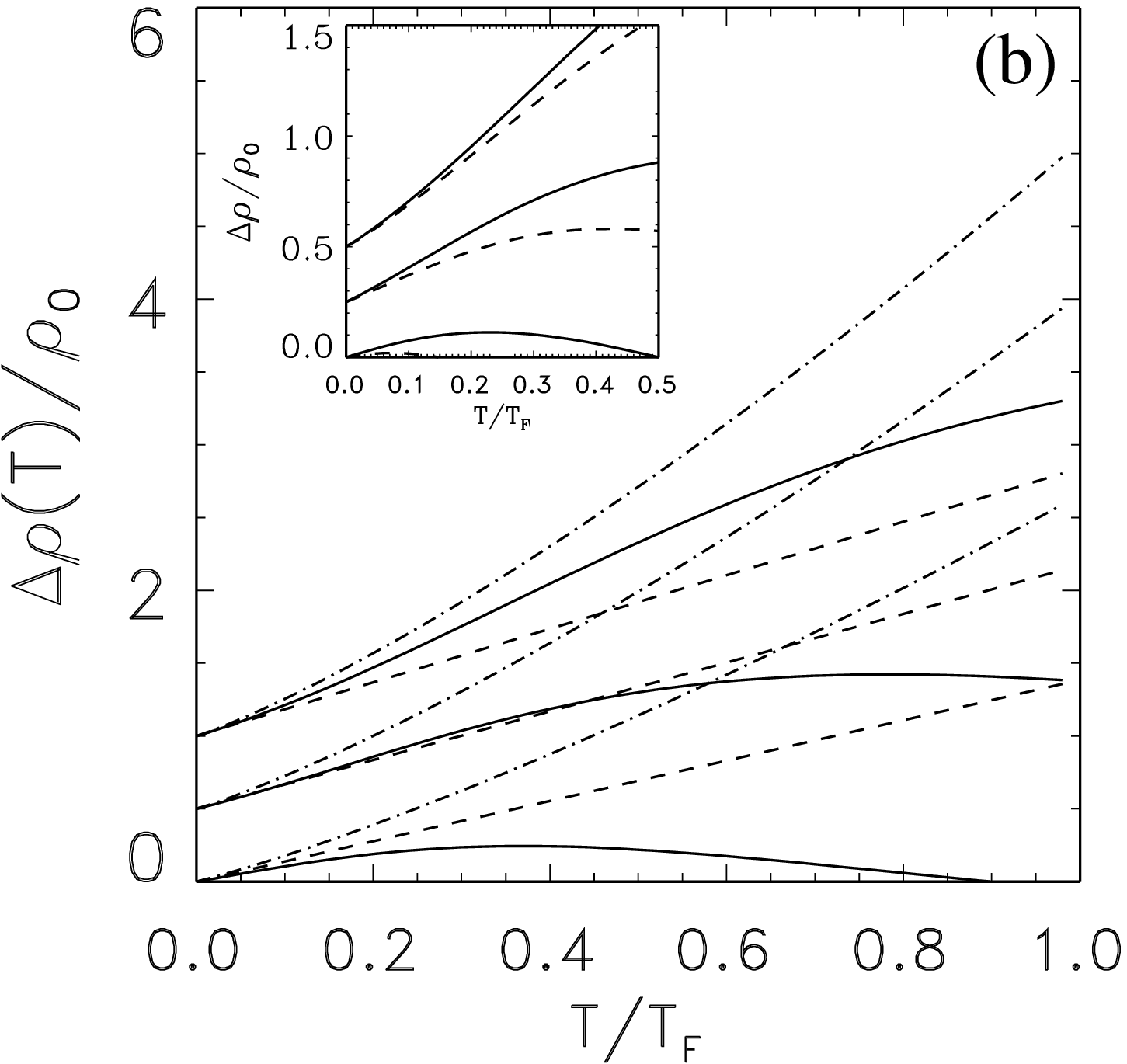}}
\vspace{0.5cm}
\caption{Calculated $\rho(T)$ as a function of $t=T/T_F$ (a) for p-GaAs with 
density $n=$21.8, 5.45, 1.36$\times 10^{10}cm^{-2}$ (from the bottom)
and (b) for n-GaAs with 
density $n=$6.1, 1.5, 0.38$\times 10^{9}cm^{-2}$ (from the bottom)
by considering only scattering from a uniform random distribution of
ionized impurities in the quantum well. 
Solid lines indicate the full RPA screening results,
dashed lines the asymptotic approximations with only 
the linear term, and dot-dashed lines with both the $O(T)$ and
$O(T^{3/2})$ terms. Offset by 0.5 for lower densities is used.
The inset in (a) shows $\rho(T)$ for charged impurities at interface. 
The inset in (b) shows $\rho(T)$ for scattering by remote charged dopants at
$d=100$\AA \ (solid lines) and $d=300$\AA \ (dashed lines)
on an expanded temperature scale. 
}
\label{fig4}
\end{figure}

To address the important question of the regime of validity of the
asymptotic linear approximation we have carried out a careful
comparison between the linear-$T$ approximation (Eqs. 1 and 2) and the
full numerical calculation of $\rho(T)$ within the RPA-Boltzmann
theory. (These results are presented in Figs.~\ref{fig3} and
\ref{fig4}.) The important
conclusion drawn from this comparison is that the regime of validity of
the asymptotic linear formula, $\triangle \rho/\rho_0 \sim 
O(T/T_F),$ is extremely restricted, and at least for the RPA screened
disorder scattering, the linear approximation holds only in the very
low temperature limit of $T/T_F < 0.05$, which, in general, is not in
the ballistic regime except at very high densities where the
metallicity is very weak.
We therefore conclude that the low-temperature asymptotic linear
regime, while being of substantial theoretical interest, is not of
much experimental relevance since $\triangle \rho \sim T$ only for
temperatures (densities) much too low (high) to be of experimental
interest. Our 
conclusion is based on a comparison of the asymptotic analytic
linear-$T$ formula with the full $T$-dependent calculation only for
the RPA screening theory (because this is the only theory where both
the asymptotic result and the full result can be calculated), but we
believe our conclusion to be quite generally valid, and even for the
interaction theory we expect the low-$T$ regime of validity of the
linear $T$ formula to be too restricted to be of much experimental
relevance.  
This is consistent with the existing experimental results where
the conductivity is  never precisely linear (at low temperatures)
essentially at any carrier densities, except perhaps at very high
carrier densities where the RPA-Boltzmann theory should


\noindent
be quantitatively valid.

In Fig.~\ref{fig3} we show our calculated $\rho(T)$ for three densities in
the (100)Si MOS 2D electron system in both the full RPA screening
theory and in the asymptotic approximations keeping only the
leading-order linear term and both the $O(T)$ and $O(T^{3/2})$ terms
in Eq. (1). We have assumed scattering from a random distribution of
charged impurities located at the interface, and subband form-factor
effects have been included in both calculations equivalently. The
inset shows $\rho(T)$ on an expanded temperature scale.

In Fig. \ref{fig4} we show similar comparisons between the calculated full
$\rho(T)$ and the asymptotic analytic approximations for 2D GaAs holes
and electrons in their experimentally relevant density regimes of
interest.  
For these two systems we show results for two different impurity
scattering mechanisms for the sake of completeness (also for
the sake of
consistency with the experimental GaAs systems where interface charged
impurity scattering is typically less important than  in Si
MOSFETs). In particular, we give results for scattering by a uniform
random distribution of (unintentional) background ionized impurities
(which are invariably present and are usually the dominant scattering
centers in the GaAs samples of the highest mobilities) and by remote
charged dopants (assumed to be randomly distributed in a 2D plane
separated a modulation distance of $d$ from the 2D electron layer),
the so-called modulation doped situation. It is obvious from Fig. \ref{fig4}
that 2D $n$-GaAs system not only has the weakest temperature
dependence but also exhibits essentially no clear linear temperature
regime. This is consistent with the very weak screening properties
(and large values of $E_F$) of 2D $n$-GaAs because of its very low
band effective mass. The real importance of the actual random impurity
distribution in the system (usually not known and has to be inferred
from a comparison of the transport data with theoretical calculations
assuming specific impurity distributions) in affecting (both
qualitatively and quantitatively) the $\rho(T)$ behavior is apparent
in the results of Fig. \ref{fig4}: The strongest temperature dependence arises
in the situation where the charged impurities are randomly distributed
in the 2D layer of the carriers and the weakest $T$-dependence arises
in the modulation doped situation (particularly for densities low
enough so that $2k_F d \ge 1$) with the interface disorder case being
intermediate. This dependence on the details of disorder is easily
understood within the screening theory by considering the role of
$2k_F$ scattering in transport: Modulation doping with $2k_F d\ge 1$
essentially completely suppresses large momentum $2k_F$ scattering
even at low temperatures because of the $e^{-2k_F d}$ term in the form
factor drastically reducing the temperature dependence due to screened
impurity scattering whereas charged impurities distributed randomly in
the 2D layer  maximizes $2k_F$ scattering for a given
system. The interaction theory of Zala {\it et al.} \cite{zala} had to make
the drastic 
approximation of a zero-range white-noise impurity disorder potential,
thus drastically (and artificially) enhancing the
$2k_F$-scattering. In real systems, other things being equal (i.e. the
mobility, the density, and the 2D system), there 


\begin{figure}
\epsfysize=3.7in
\centerline{\epsffile{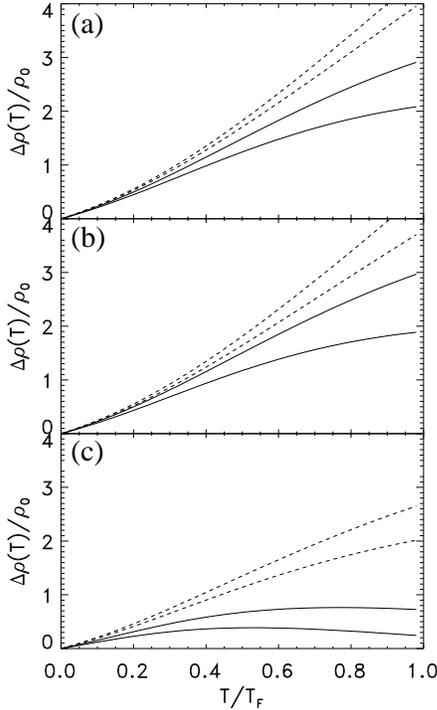}}
\caption{The calculated pure-2D (dashed lines) and {\it quasi}-2D
  (solid lines)
temperature-dependent resistivity for screened bulk charge impurity
scattering (a) for Si-MOSFET with $n=$ 10, 20$\times
10^{10}cm^{-2}$, (b) for p-GaAs with $p=$ 2.0, 5.0$\times
10^{10}cm^{-2}$, and (c) for n-GaAs with $n=$ 0.5, 1.0$\times
10^{10}cm^{-2}$. The lower lines represent higher densities.
}
\label{fig5}
\end{figure}
\noindent
would be a strong
dependence of the detailed behavior of $\rho(T)$ on the actual random
impurity distribution in the system through the form factor effect. 
(Note that the low-T asymptotic linear formula does not depend on the
impurity distribution or on the range of disorder scattering, but its
temperature regime of validity may very well depend on the nature of
scattering in the 2D system.)


\section{solid state physics effects: quasi-2D layer width and bare
  impurity potential range}

We now discuss the quantitative significance of various
solid state physics  effects on the 2D metallicity. The specific
effects we discuss are the {\it quasi}-2D nature of the semiconductor
layers under consideration and the nature of bare impurity disorder
({\it i.e.} long range versus short range) in determining the
temperature dependence of 2D resistivity. (Both of these effects are
ignored in Ref. \onlinecite{zala}. The inclusion of the quasi-2D
form-factor effect in the interaction theory is straightforward, but
the inclusion of long-ranged bare disorder, e.g. charged impurity
scattering, in the interaction theory is nontrivially difficult.) In
Fig. \ref{fig5} we compare the calculated 2D and {\it quasi}-2D
temperature-dependent resistivity for screened bulk charge impurity
scattering, finding that the metal-

\begin{figure}
\epsfysize=3.5in
\centerline{\epsffile{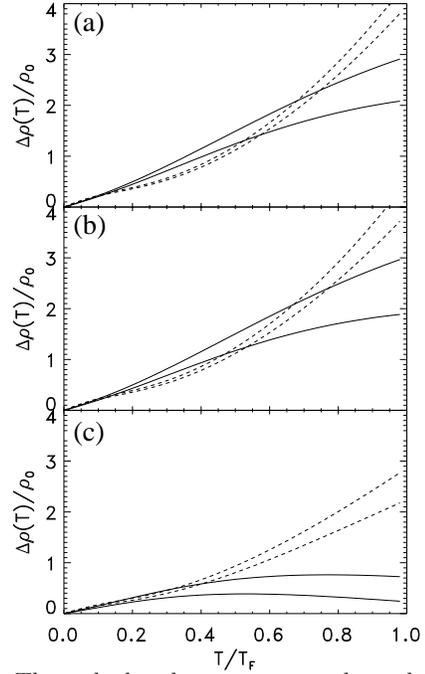}}
\caption{The calculated temperature dependent resitivities for
realistic long-range charged impurity scattering (solid lines) 
and for hypothetical
short-range delta-function impurity scattering (dashed lines)
(a) for Si-MOSFET with $n=$ 10, 20$\times
10^{10}cm^{-2}$, (b) for p-GaAs with $p=$ 2.0, 5.0$\times
10^{10}cm^{-2}$, and (c) for n-GaAs with $n=$ 0.5, 1.0$\times
10^{10}cm^{-2}$. The lower lines represent higher densities.
}
\label{fig6}
\end{figure}
\noindent
licity is substantially
overestimated in the strictly two-dimensional approximation. Thus, it
is essential to include the quasi-2D nature of the semiconductor
system in order to obtain quantitatively accurate temperature
dependence although the 2D approximation is qualitatively correct.

In Fig. \ref{fig6} we show a comparison between the results calculated
for realistic long-range charged impurity scattering and hypothetical
short-range delta-function (in real space) impurity scattering (both
equivalently screened by the finite temperature RPA dielectric
function of the 2D carriers). We point out that the dominant disorder
in 2D semiconductor systems arises from long-range (Coulomb) charged
impurity scattering. In Fig. \ref{fig6}, the calculated temperature
dependence 
is obviously substantially stronger for short-range bare disorder. It
is important to emphasize in this context that the leading-order
asymptotic dependence of $\triangle \rho/\rho_0$ on $T/T_F$ is {\it
independent} of the range of the bare disorder since it depends (see
Eqs. 1 and 2) on the electron-impurity interaction only through
$V_{\rm e-i}(2k_F)$, {\it i.e.} through the constant (momentum space)
impurity potential defined at the wave vector $q=2k_F$, making long-
and short-range bare disorder completely equivalent for the low
temperature asymptotic temperature dependence. On the other hand, the
full temperature dependence depicted in our Fig. \ref{fig6} manifestly
demonstrates that, except at the lowest values of $t=T/T_F \le 0.1$,
the actual temperature dependent resistivity depends quite
significantly on the range of bare disorder with short-range bare
disorder providing substantially stronger temperature dependence than
the long-range Coulomb disorder due to random charged impurity
scattering. Since we have already argued in Sec. III that the
asymptotic linear temperature dependence does not really apply, except
at extremely low temperatures (or high carrier densities), the range
of impurity disorder potential takes on special significance in the
theoretical analyses.


\section{comparison with experiment}

Although the primary goal of this article is to establish and clarify
certain theoretical principles (e.g. the validity of the asymptotic
low temperature expansion, the strength of metallicity as reflected in
the $q_0$, $t_0$ scaling behavior, the importance of various realistic
solid state physics effects such as the long-range versus the
short-range nature of the bare impurity disorder potential or the
quasi-2D nature of the semiconductor systems of experimental interest)
within the RPA-Boltzmann theory of 2D carrier transport in
semiconductor systems, it is important to ask about the empirical
validity of the zeroth order RPA-Boltzmann theory in the context of
the large amount of the available temperature dependent 2D transport
experimental data. We have in fact carried out a number of comparisons
between our theory and the experimentally measured temperature
dependent resistivity in several different 2D systems of current
interest \cite{DH1,DH2,senz,lewalle,lilly,noh1,hwang03,DH11,DH12}. It
is important in this context to remember that such comparisons between
experiment and theory is necessarily qualitative in nature since the
experimental data, even in the same material system, show strong
sample to sample variations, and the measured $\rho(T,n)$ in
different Si MOS systems (or for that matter, in different n-GaAs or
p-GaAs systems) often have significantly different quantitative (and
sometimes even qualitative) dependence on temperature and carrier
density. This is understandable since different samples in general may
have significantly different bare disorder potential and system
parameters, and therefore a universal quantitative behavior of
$\rho(T,n)$ cannot be expected (since there is no universal
quantitative behavior for the experimental data themselves).

The important question therefore is the extent to which a particular
theory explains the qualitative behavior of the observed $\rho(T,n)$ in
2D systems. The RPA-Boltzmann theory discussed in this paper is unique
in this respect since it is the {\it only} theory which is capable of
producing the full temperature and density dependent 2D resistivity
which can, in principle, be compared with the experimental data. The
theory is still not uniquely defined since all the system parameters
(e.g. effective 

\begin{figure}
\epsfysize=2.8in
\centerline{\epsffile{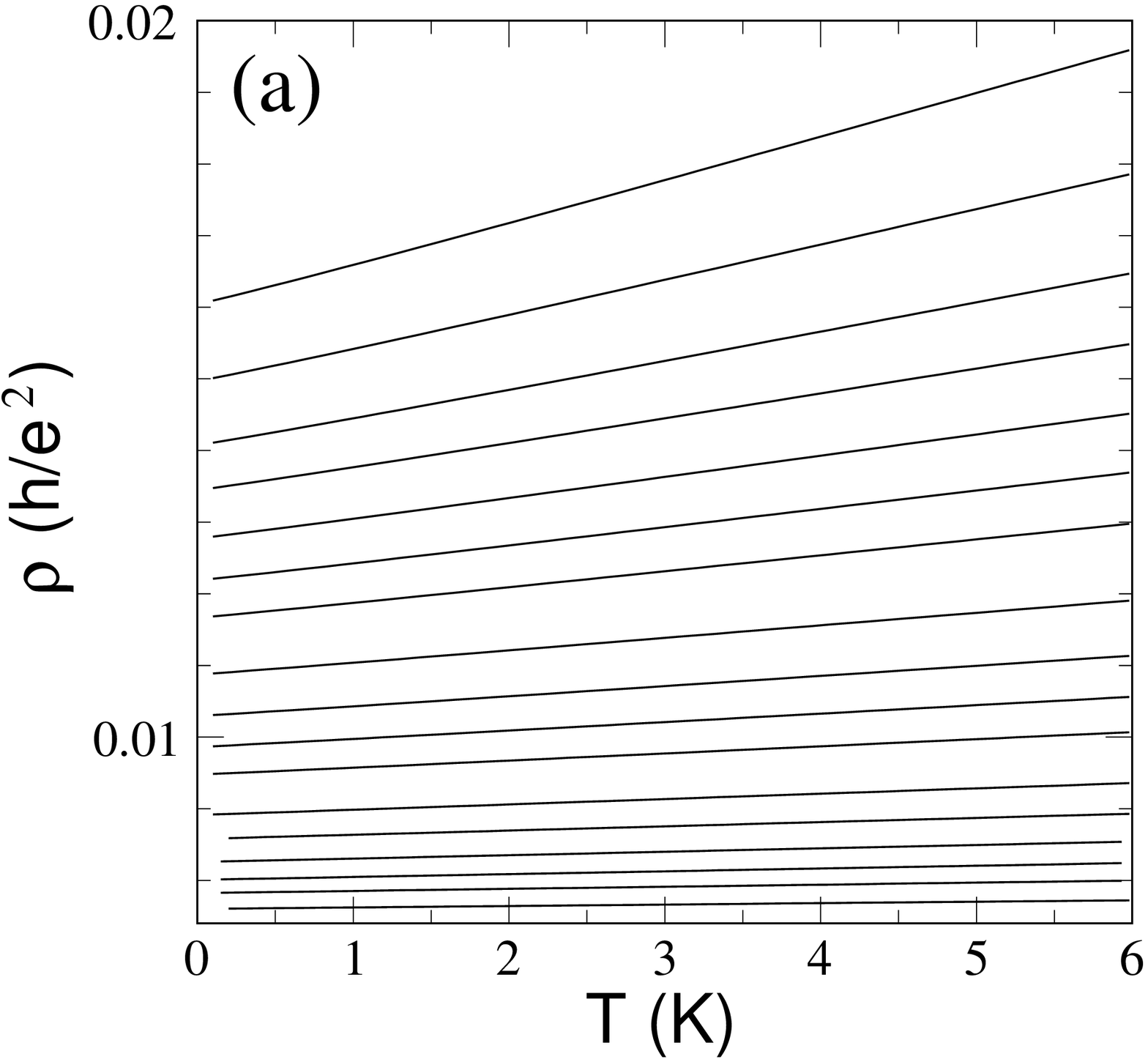}}
\vspace{0.5cm}
\epsfysize=2.8in
\centerline{\epsffile{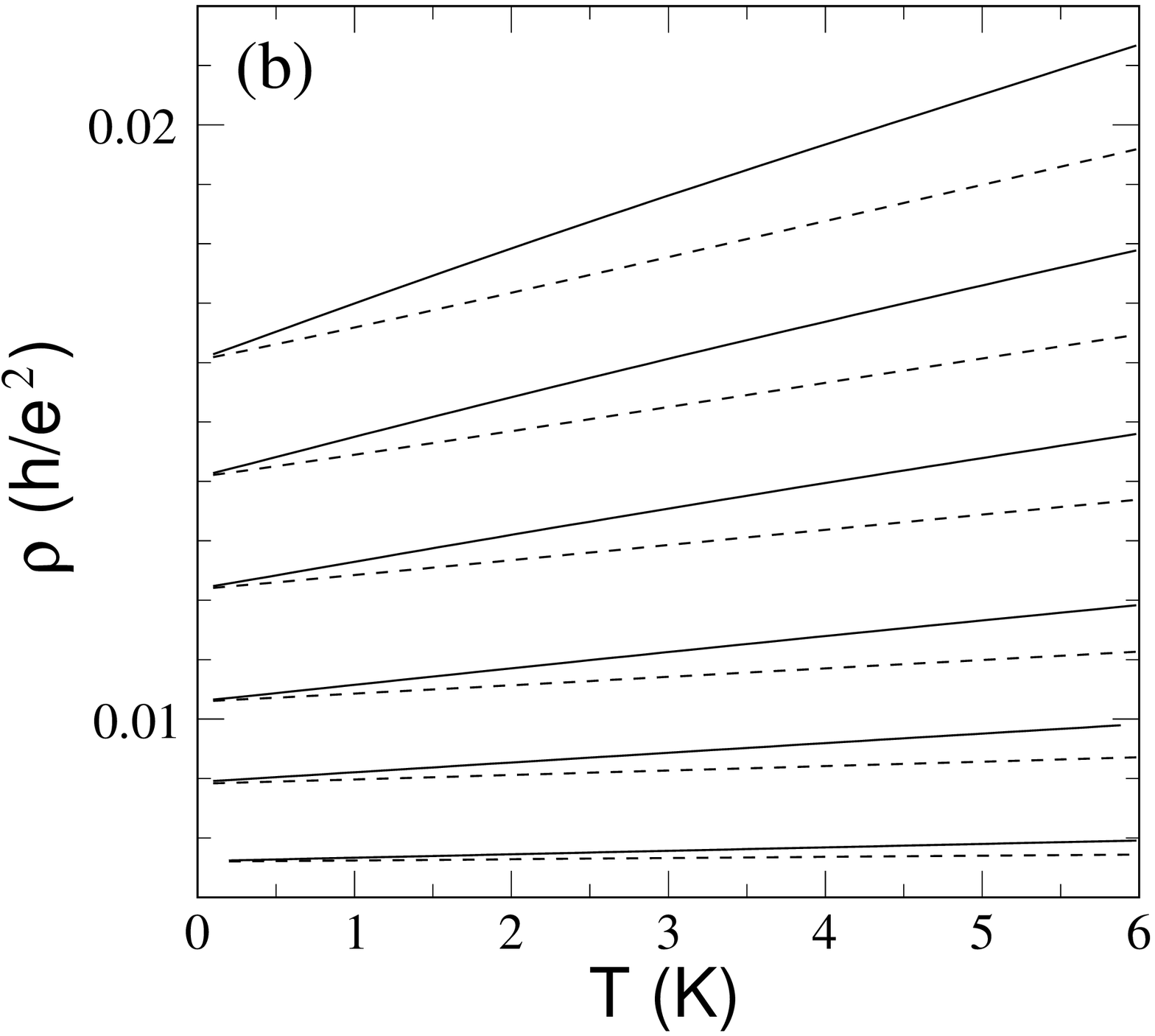}}
\vspace{0.5cm}
\caption{
(a) Calculated $\rho(T,n)$ 
within the RPA-Boltzmann transport theory
assuming only interface 
charged impurity scattering, for all the densities
corresponding to Fig. 1(a) (Si-22 sample) in ref. 22. 
(b) Calculated $\rho(T,n)$ 
with long-range  Coulomb impurities randomly
distributed at the interface (dashed lines) and the ideal
zero-thickness 2D layer approximation with zero-range $\delta$-function
potential bare impurity scatterers (solid lines)
for six
representative carrier densities $n=$ 5.7, 6.9, 8.7, 11.7, 16.5,
35.7$\times 10^{11} cm^{-2}$ (top to bottom).
}
\label{fig7}
\end{figure}

\noindent
mass, depletion charge density, the precise metallic
carrier density, etc.) are never really accurately known, and more
importantly, the detailed quantitative parameterization of the bare
disorder potential is never available for any sample. In general,
there could be many independent sources for the bare disorder
(e.g. bulk and interface charged impurities, surface roughness, remote
impurities, intervalley scattering, phonons, alloy scattering), and
with a sufficient number of adjustable free parameters characterizing
different kinds of disorder assumed to exist in a system one may very
well be able to obtain essentially complete (but, meaningless)
agreement with experimental data for any particular sample.

With all these caveats in mind we carry out a comparison between our
theory and a recent set of experimental data \cite{pudalov} for
$\rho(T,n)$ in a Si MOS system. The specific sample we choose for this
comparison is the Si-22 sample (see Fig. 1(a) of
ref. \onlinecite{pudalov}) with a quoted ``peak mobility'' of
33,000 $cm^2/Vs$. In ref. \onlinecite{pudalov} a series of $\rho(T)$
curves, for $T \approx 0.5K - 5.0K$, are presented for this sample in
the 2D carrier density range of $n=5.7
- 35.7\times 10^{11} cm^{-2} $ with the typical resistivity values
spanning between 0.002 $h/e^2$ to 0.02 $h/e^2$. The temperature induced
fractional change in $\rho(T)$ in this $0.5-5.0K$ temperature window
spans between a few percent at higher densities to about 35\% 
at the lowest densities. Visually the $\rho(T)$ curves of Fig. 1(a) in
ref. \onlinecite{pudalov} corresponding to the Si-22 sample all seem
to obey the asymptotic linear temperature dependence although, in
general, there is significant deviation from linearity {\it both} at
the lowest ($\le 0.5K$) and at the highest ($\le 5K$) temperatures.

\begin{figure}
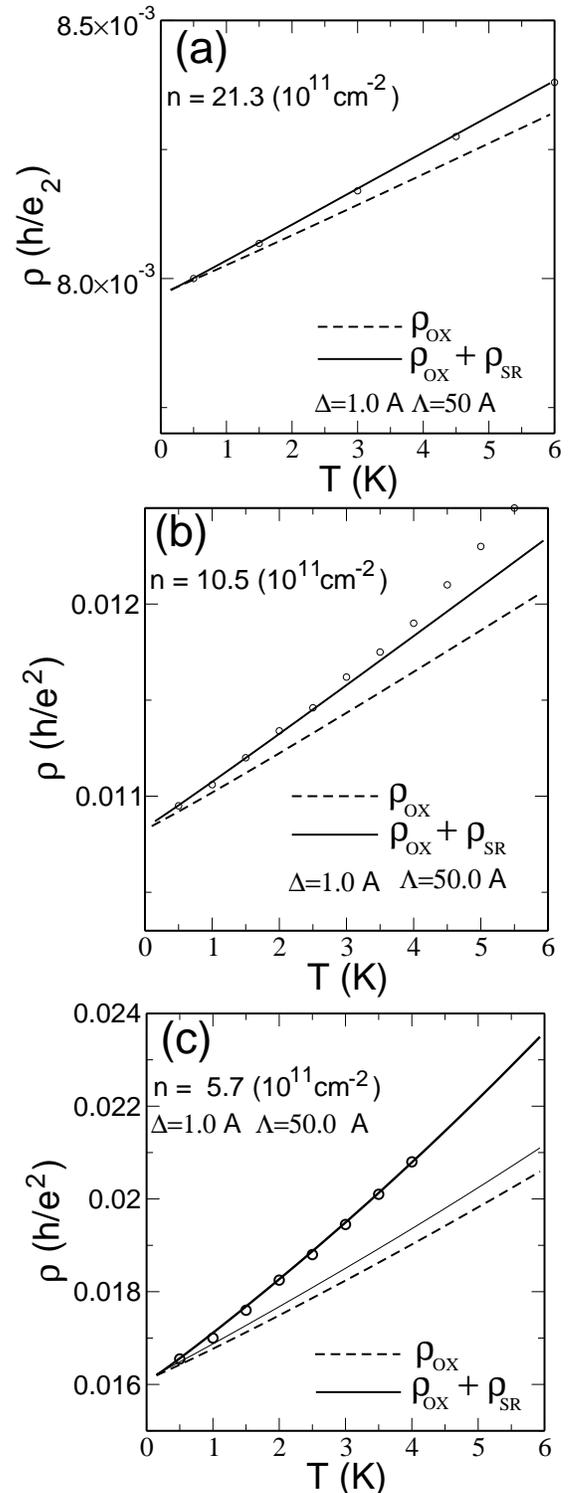

\epsfysize=2.6in
\centerline{\epsffile{fig8a.eps}}
\epsfysize=2.6in
\centerline{\epsffile{fig8b.eps}}
\epsfysize=2.6in
\centerline{\epsffile{fig8c.eps}}
\caption{
Calculated $\rho(T)$ for three  carrier densities
(a) $n=$ 21.3$\times 10^{11}cm^{-2}$,
(b) $n=$ 10.5$\times 10^{11}cm^{-2}$,
(c) $n=$ 5.7$\times 10^{11}cm^{-2}$,
including realistic surface roughness and
interface (oxide) charged impurity scattering.
In (c) the thin (thick) solid line represents the result using the band
effective mass (the measured effective mass from Ref. [24]).
Here $\Delta$ is the average displacement of the interface and
$\Lambda$ is of the order of the range of its spatial variation in the
direction parallel to the surface.
}
\label{fig8}
\end{figure}

In Figs. \ref{fig7} and \ref{fig8} we show our calculated $\rho(T,n)$
for the Si-22 sample \cite{pudalov} within the RPA-Boltzmann transport
theory assuming screened disorder scattering. In Fig. \ref{fig7}(a) we
show the 
calculated RPA-Boltzmann resistivity, assuming only interface (or
oxide) charged impurity scattering, for all the densities
corresponding to Fig. 1(a) in ref. \onlinecite{pudalov}. The results
shown here look qualitatively similar to the experimental data
[cf. Fig 1(a) in Ref. \onlinecite{pudalov}] except that the
temperature dependence is somewhat weaker in the theory. In
Fig. \ref{fig7}(b) we show 
the theoretical results, $\rho(T)$ with $T=0.25-6K$, for six
representative carrier densities $n=$ 5.7, 6.9, 8.7, 11.7, 16.5,
35.7$\times 10^{11} cm^{-2}$ (top to bottom in Fig. \ref{fig7}) using
impurity 
scattering as the only resistive mechanism in the system, comparing
the quasi 2D realistic situation and a
long-range bare impurity potential with the idealized 2D situation
and a short-range bare impurity potential. 
The two sets of calculated results in Fig. \ref{fig7} corresponding to
the realistic quasi-2D system 
(with finite 2D layer thickness, semiconductor-insulator dielectric
mismatch, etc.) with long-range bare Coulomb impurities randomly
distributed at the interface (dashed lines) and the ideal
zero-thickness 2D layer approximation with zero-range $\delta$-function
potential {\it bare} impurity scatterers (the model, for example, of
ref. \onlinecite{zala}) providing the resistive scattering mechanism
(solid lines). We note several important features of the results
presented in Fig. \ref{fig7}: (1) The theoretical results for
$\rho(T)$ appear 
to be approximately linear in $T$ for all the carrier densities, being
qualitatively very similar to the experimental data shown in Fig. 1(a)
of ref. \onlinecite{pudalov}; (2) in spite of this qualitative
linearity of $\rho(T)$, the actual curves are nonlinear
except at the highest densities; (3) an approximate measure (actually a
lower bound) of this nonlinearity is the difference between the 


\noindent
solid
and dashed lines in Fig. \ref{fig7} since the slopes of the two lines
are the 
{\it same} in the $T \rightarrow 0$ limit where both are strictly 
linear in $T/T_F$ as discussed above in section III; (4) the
calculated results, while being qualitatively similar to the data
\cite{pudalov}, disagree quantitatively with experiment with the
quantitative disagreement increasing with decreasing carrier density;
(5) the {\it unrealistic} strictly 2D approximation with zero-range
$\delta$-function {\it bare} disorder potential (solid lines in
Fig. \ref{fig7}) actually agrees much {\it better} with the
experimental data 
than the realistic quasi-2D calculation with charged Coulomb impurity
scattering as the bare disorder (dashed lines) since the experimental
data of ref. \onlinecite{pudalov} {\it shows} stronger temperature
dependence quantitatively than the theoretical results of Fig. \ref{fig7}
(again this is consistent with the discussion of Sec. IV above where
$\delta$-function bare impurity disorder and/or strict 2D
approximation turns out to give much stronger temperature dependence
than the realistic quasi-2D system with charged Coulomb scattering)
--- we emphasize, however, that both the solid and the dashed lines
have the same slope in the $T/T_F \rightarrow 0$ limit reinforcing the
lack of linearity in the $\rho(T)$ behavior.

To go beyond the qualitative agreement with experiment depicted in
Fig. \ref{fig7}, we consider in Fig. \ref{fig8} a {\it more realistic}
situation taking 
into account, in addition to the charged Coulombic impurity disorder,
the surface roughness scattering by the Si-SiO$_2$ interface which has
been known for a very long time \cite{Ando_rmp} to be the
dominant scattering mechanism affecting carrier transport in Si
MOSFETs at relatively ``higher'' carrier densities whence the 2D
electron gas resides rather close to the Si-SiO$_2$ interface making
the roughness scattering to be significant. We include the surface
roughness scattering in our theory in the standard manner
\cite{Ando_rmp}, screening it within the RPA theory similar to the
charged impurity scattering theory \cite{stern}. In
Figs. \ref{fig8}(a)-(c) we 
show our calculated $\rho(T)$ for three representative carrier densities
(all for the Si-22 sample of ref. \onlinecite{pudalov}) $n=$ 5.7, 10.5,
21.3$\times 10^{11}cm^{-2}$ including realistic surface roughness and
interface (oxide) charged impurity scattering. 
At higher densities the
quantitative agreement between the realistic calculations (solid
lines) including both charged impurity and surface roughness
scattering and the experimental data is very good whereas at the lower
density (5.7$\times 10^{11}cm^{-2}$) the agreement is systematically
poorer at higher temperatures where the theory consistently
underestimates the experimental temperature dependence. 
However,  if we use the actual density dependent effective mass
measured in the experiment rather than the band mass, 
we obtain much better agreement between our result and the experimental
data at low densities. In Fig. \ref{fig8}(c) we show our calculated
resistivity (thick solid line) using the measured effective mass taken
from ref. \onlinecite{pudalov1}. 
We note, however, that the effective mass renormalization (and its
effect on transport) is a rather subtle issue, and, in spite of the
excellent quantitative agreement obtained in Fig. \ref{fig8}(c) using the
experimentally measured density dependent effective mass, it is
unclear whether the band mass or the renormalized mass should be used
in our RPA-Boltzmann transport theory.

The quantitative agreement shown in Fig. \ref{fig8} can be further
improved by 
including {\it bulk} charged impurity scattering due to random
ionized impurity centers in the semiconductor material itself, whose
relative importance increases at lower densities as the 2D electrons
reside on the average more deeply inside Si (and away from the
interface). If we include in addition scattering by {\it neutral}
impurities (with short-range bare impurity potential in contrast to
the long-range bare impurity potential arising from the charged
impurities), we can essentially obtain quantitative agreement
between theory and experiment by suitably adjusting the relative
impurity densities among oxide charged impurities, bulk ionized
impurities, and neutral short-range impurities (in addition, of course,
to short-range surface roughness scattering). But such a parametrized
quantitative agreement between theory and experiment is essentially
a device simulation exercise and is completely meaningless 
from the perspective of the fundamental physics of 2D carrier systems,
since it
does not tell us anything more than what the results of
Figs. \ref{fig7} and \ref{fig8} 
already tell us. The important point to realize here is that even the
qualitatively valid results of Fig. \ref{fig7} already show the same
level of 
{\it quantitative} agreement between the RPA-Boltzmann  theory and the
Si-22 data of ref. \onlinecite{pudalov} as the existing agreement
among the experimental data from different Si MOSFET samples at same
densities and temperatures --- this can be easily seen just by
comparing to the data in Fig. 1(a) of ref. \onlinecite{pudalov}
corresponding to the Si-22 sample with those in Fig. 1(b), (c) of
ref. \onlinecite{pudalov} corresponding to the Si-15 sample (in fact,
sample to sample variations in the observed $\rho(T,n)$ in various Si
MOSFETs are typically much larger than the quantitative agreement we
find without adjusting any parameters in Fig. \ref{fig8}). Our results
shown in 
Figs. \ref{fig7} and \ref{fig8} also reinforce the real danger of
insisting just on 
``explaining'' theoretically the low-temperature slope of the
resistivity (which automatically assumes a linear temperature
dependence in both resistivity and conductivity) as has been
fashionable in the recent experimental work \cite{noh,pudalov} motivated
by the asymptotic low temperature analysis of the interaction
theory \cite{zala}. As Figs. \ref{fig7} and \ref{fig8} explicitly
demonstrate, an agreement between 
theory and experiment on $d\rho/dT$ in the $T\rightarrow 0$ limit is
absolutely no guarantee for a quantitative agreement between theory
and experiment in a reasonable temperature range. Finally, we point
out that the unrealistic ideal 2D approximation and bare zero-range
disorder potential misleadingly provides ``better'' quantitative
agreement between theory and experiment, which of course does not mean
anything about the physics of 2D carrier systems.
We also note in this context that the quantitative agreement between
the RPA-Boltzmann theory and the low density experimental data can be
substantially improved \cite{DH1} by assuming an effective lower
density for the {\it free} carriers (participating in ``metallic''
transport) than the nominal 2D carrier density, which is equivalent to
invoking a lower effective Fermi temperature for the system (this
could also arise from an enhanced effective mass at lower densities as
has been experimentally observed), thus effectively enhancing the 
theoretical temperature dependence.

It is instructive to separate on the possible reasons for the
systematic deviation of the experimental $\rho(T)$ from the
RPA-Boltzmann theory at lower densities ($\le 5\times 10^{11}
cm^{-2}$) and moderate temperatures ($\ge 2$K). There are two general
possibilities: limitation or shortcomings of the RPA-Boltzmann
theoretical scheme and the participation of scattering mechanisms left
out of our model. The RPA-Boltzmann transport theory is obviously
quite approximate since it is fundamentally semiclassical in nature
leaving out all interaction, localization, and quantum interference
effects. It is, however, difficult to understand why these effects
(i.e., localization, interaction, interference), which are left out of
the RPA-Boltzmann transport theory, would become quantitatively more
important at higher temperatures. In fact, quantum effects should be
progressively weaker as $T/T_F$ increases, and therefore the
systematic underestimation of the experimental $\rho(T)$ in our
RPA-Boltzmann theory is very unlikely to arise from the theoretical
approximations of our model. We believe that the systematic apparent
disagreement between experiment and theory at higher temperatures and
lower densities in Fig. \ref{fig8}, if real, must arise from a missing
scattering mechanism (neglected in our model) which takes on
significance at higher temperatures. Such additional (high
temperature) scattering may be due to intervalley scattering in Si
and/or surface/oxide phonon modes not included in our theory.

In discussing the theoretical approximations of our model it is
important to emphasize, particularly in the context of the results
shown in Figs. \ref{fig7} and \ref{fig8} for comparison with the data of
Ref. \onlinecite{pudalov}, that the leading-order single-site
scattering approximation used in our theory is quite accurate since
the typical values of `$k_Fl$' (where $k_F$ and $l$ are respectively
the Fermi wave vector and the carrier mean free path) are rather large
for the results depicted in Figs. \ref{fig7} and \ref{fig8}. It is
easy to show that for 
a 2D system $k_Fl = \tilde{\sigma}/g_{\nu}$, where $\tilde{\sigma}
\equiv \sigma/(e^2/h)$ is the dimensionless conductance of the 2D
system. For the results of Ref. \onlinecite{pudalov}, as shown in
Figs. \ref{fig7} and \ref{fig8} of our paper, $k_Fl$ is typically 100
or higher, making 
the single-site approximation well-valid. Thus, we do not believe that
higher-order impurity scattering effects (or localization effects) are
important in the regime of density and temperature covered in our
Figs. \ref{fig7} and \ref{fig8}. Obviously, at lower densities, near
the critical 
density for the metal-insulator transition (where $k_Fl \le 1$),
higher-order impurity scattering effects become very important, and
our simple RPA-Boltzmann theory breaks down.

\section{discussion and conclusion}

We have carried out a critical comparison between the 
full $\rho(T)$ calculated within the RPA-Boltzmann theory and its 
asymptotic ($T/T_F \rightarrow 0$) linear-$T$ approximation, finding that 
the linear approximation strictly applies only at very low temperatures, 
typically $T/T_F < 0.05$, at least within the RPA screening theory.
In addition, we have provided a qualitative 
explanation, based on our approximate finding 
$\triangle \rho(T,n) \approx \triangle \rho(t,q_0)$ where $t \equiv 
T/T_F$ and $q_0 \equiv q_{TF}/2k_F$, for the relative strength of 
metallicity in various 2D systems (showing in the process that the 
dimensionless screening parameter $q_0$, and {\it not} the dimensionless 
interaction parameter $r_s$,
provides a better zeroth-order qualitative description for the metallicity 
trend in various materials).
We have also carried out a detailed analysis of 
$\rho(T,n)$ in the strictly 2D limit ({\it i.e.}, with the subband form 
factor $f\equiv 1$) and 
using short-range bare disorder.
The strictly 2D results are in general 
quantitatively incorrect showing a more prominent linear low-$T$ regime 
and manifesting much stronger 
metallicity than the realistic quasi-2D results.
Similarly, the short-range bare disorder manifests stronger
temperature dependence than the realistic charged impurity scattering
case although they both have the same leading-order temperature
dependence. We have also carried out our theoretical transport
calculations (not presented in this paper) including local field
corrections to 2D screening by 
going beyond RPA (which is exact only in the high density or in the
high temperature limit). Our results 
with local field corrections are qualitatively very similar to the RPA 
screening results presented in this paper except that the temperature 
dependence of $\rho(T)$ is in general somewhat weaker. 
We restrict ourselves to presenting only RPA screening results because
there is no unanimity in the literature about the best possible local
field corrections and also because the temperature dependence of local
field effects are in general unknown.
We point out that recent experiments attempting
to verify the interaction theory \cite{noh,pudalov,noh1}
have produced conflicting 
results mainly because a clear linear-$T$ 
regime in conductivity satisfying $T_D \ll T \ll T_F$ seems {\it not} 
to exist in low density ``metallic'' 2D systems (as we explicitly show 
for the screening theory results in this paper) in any experimentally 
accessible range of the low-$T$ data.
One should therefore be extremely careful in applying any
leading-order asymptotic temperature expansion to the 2D resistivity
in analyzing experimental data.

This caution in comparing 2D transport data with the interaction
theory \cite{zala} is particularly warranted in view of the minimal
necessary (but by no means sufficient) conditions that must be
satisfied for such a comparison to be meaningful: (1) The `weak'
temperature constraint $T_D \ll T \ll T_F$ must be obeyed so as to be
in the low temperature ballistic limit and phonon scattering must be
negligible, (2) a clear linear temperature dependence in the low
temperature conductivity ({\it not} resistivity) $\sigma(T) =
\sigma_0[1 + C_1(T/T_F)]$, where the slope $C_1$ depends on density,
must be observed over a {\it reasonable} (a decade or more) range of
temperature satisfying the ballistic constraint $T_D \ll T \ll T_F$;
(3) the actual temperature dependent conductivity correction, $\Delta
\sigma /\sigma_0 = | \sigma(T) -\sigma_0|/\sigma_0$, must be very
small ($\Delta \sigma/\sigma_0 \ll 1$) for the leading-order
interaction theory to be applicable. This set of necessary conditions
is sufficiently restrictive so that no 2D transport experiment
actually satisfies all three conditions except at very high carrier
densities where the RPA-Boltzmann theory gives quantitatively accurate
results \cite{DH12}.

Before concluding we discuss the experimental relevance of our
($q_0$, $t$) scaling prediction and the theoretical relevance (or
validity) of our RPA screening approximation. A cursory examination of
the available 2D transport data shows that our predicted scaling of
$\Delta \rho/\rho_0$ with the parameters $q_0=q_{TF}/2k_F$ and
$t=T/T_F$ works on a qualitative level 
as a zeroth order description for GaAs
electrons and holes, and also for electrons in Si MOS but only at low
values of $T/T_F$ ($\le 0.2$). At higher temperatures phonon
scattering becomes significant \cite{DH2,lilly,noh1}
in 2D GaAs systems whereas in Si MOS structures, where phonon effects
should be negligible, the temperature dependence of resistivity at
lower metallic densities becomes stronger than our prediction for
reasons unclear to us. 
We note that this systematic under-estimation of experimental
temperature dependence of $\rho(T,n)$ at lower (higher) values of $n$
($T$) in the RPA scattering theory is unlikely to be arising from the
interaction effects considered in Ref. \onlinecite{zala} since the
disagreement arises in the non-asymptotic regime of $T/T_F \ge 0.2$
where the experimental $\rho(T,n)$ is manifestly non-linear in $T/T_F$
making the theory of Ref. \onlinecite{zala} inapplicable.
We can speculate several possible reasons for
this unusually strong temperature dependence of Si MOSFETs: (1)
Somehow the effective Fermi temperature (and the Fermi wave vector) in
the system could be smaller than the nominal Fermi temperature (or the
Fermi wave vector) obtained on the basis of (100) Si inversion layer
band mass and measured carrier density (leading to enhanced values of
$q_0$ and/or $t$) -- for example, the effective mass 
could be larger due to renormalization or the effective free carrier
density smaller (e.g. due to trapping by interface defects); (2) there
could be additional scattering mechanisms 
(e.g. intersubband scattering between
different valleys in Si, which 
would be enhanced at higher temperatures) not included in our theory;
(3) RPA could be failing systematically at higher values of $T/T_F$
(this is an unlikely scenario since RPA should become a better
approximation at higher $T/T_F$ values, and going beyond RPA including
local field corrections does not help in this respect).

We note that we have ignored in this work any damping correction to
the RPA screening function arising from the impurity scattering
effect. In particular, for $T \ll T_D$ the temperature-induced thermal
suppression of $2k_F$ screening, which is crucial in producing the
strong temperature dependence of resistivity \cite{stern} at low
temperature, becomes completely ineffective since collisional
broadening or damping effects induced by impurity scattering already
suppresses screening at $q=2k_F$. Therefore, for $T\ll T_D$ we expect
the 2D resistivity $\rho(T)$ to become essentially independent (or a
very weak sublinear function) of temperature. This is, in fact,
precisely the experimental observation --- the strong (and often
approximately linear) temperature dependence of the low temperature
resistivity almost always shows a saturation at very low temperatures
for $T \ll T_D$. This damping or broadening induced screening
suppression for $T \ll T_D$ has a more subtle effect also. At low
carrier densities $T_F$ is low and $T_D$ is typically high since
scattering effect is strong at low density; therefore at very low
densities, a ``metallic'' 2D system may not manifest strong
``metallicity'' because the damping induced low temperature saturation
of $\rho(T)$ will be more important as $T_D$ approaches $T_F$ at low
densities. This effect is also consistent with experimental
observations. We have discussed elsewhere \cite{DH1} the damping
correction to screening in some details in the context of the 2D MIT
phenomenon. In this article we have restricted ourselves mostly to
discussing the ``ballistic'' $T \gg T_D$ regime of 2D metallicity, and
as such, we have decided to ignore the damping correction. The other
reason for ignoring the damping correction is that the precise Dingle
temperature value $T_D$ to be used in the theory is unknown, and
therefore it introduces an unknown free parameter which we wish to
avoid. Also, our purpose of comparison with the interaction theory of
Zala {\it et al}. \cite{zala} is not well-served by having the damping
correction since the interaction theory has been explicitly developed
for the ballistic regime. We do mention, however, that introducing a
Dingle temperature induced damping correction to the RPA screening
function will, in general, reduce the overall temperature dependence
of our calculated resistivity with the low-temperature ($T \ll T_D$)
resistivity showing an approximate saturation behavior \cite{DH1}.

We emphasize
that the screening theory produces excellent qualitative 
agreement with the existing experimental data and
provides a good zeroth order ($q_0$, $t$) scaling description, which
is all one should expect from the simple Drude-Boltzmann RPA-screening
theory for a strong-coupling (i.e. low carrier density) problem. The
question of how valid Boltzmann theory is for understanding 2D
``metallicity'' is a difficult question to answer. A serious problem
in this respect is the fact that this is the {\it only} quantitative
theory that exists in the literature for studying 2D metallicity, and
therefore its validity can only be judged {\it a posteriori} through
an empirical comparison with the experimental data. (There is simply no
competing alternative quantitative theory in this problem, making the
discussion of the validity of the RPA-Boltzmann theory somewhat
meaningless -- we emphasize in this context that the interaction
theory of ref. \onlinecite{zala} is {\it not} an alternative theory,
it is an extension of the RPA theory to incorporate higher-order
quantum interaction corrections within necessarily a highly
restrictive model; the interaction theory applies only when the
temperature correction to conductivity is very small and linear.) We
believe that the RPA screening description 
should be qualitatively well valid in this problem as long as the
dominant disorder in the 2D semiconductor systems arises from
long-ranged random charged impurity scattering (as is the case here at
low carrier densities). This is because RPA is a physically motivated
approximation, leading to the screening of the long-ranged Coulomb
scattering potential to a short-ranged screened disorder. The
approximation becomes exact only in the limit of high density and high
temperature, but the 2D metallicity manifests itself only for finite
values of $T/T_F$ (i.e. for $T/T_F$ extremely small $\rho$ does {\it
  not} show strong $T$-dependence!) and as such, RPA should be a
reasonable description. We believe that the exactness of RPA in the
$r_s \rightarrow 0$ limit has been overemphasized in the literature
-- RPA remains a qualitatively accurate description of metallic
systems even for 
large values of $r_s$ (particularly at finite $T/T_F$) as
long as there is no quantum phase transition to a non-Fermi liquid
phase. The 2D metal-insulator-transition (2D MIT) being essentially a
``high-temperature'' phenomenon, RPA, in our opinion, is a reasonable
description. The fact that our calculated transport results change
little by including local field
corrections (beyond RPA) further reinforces
the qualitative validity of RPA to this problem. It should, however,
be emphasized that like any other (nonperturbative) uncontrolled
approximation (e.g. DMFT, LDA) the quantitative accuracy of RPA is
difficult to ascertain from a purely theoretical viewpoint.
One advantage of the RPA-Boltzmann minimal transport model adopted in
our work is that transport calculations can be carried out for
arbitrary temperatures and carrier densities with concrete comparisons
with experimental data. We have carried out one such comparison with a
recent experiment \cite{pudalov} in this paper (sec. V), and earlier
comparisons exist in the literature
\cite{senz,lewalle,lilly,stern,noh1,DH11,DH12}. The general conclusion
one can draw from these comparisons is that the RPA-Boltzmann theory
is a reasonably successful {\it zeroth order} model for 2D transport
properties, providing a qualitative (and intuitively appealing)
explanation for the strong 2D metallicity at low carrier densities. As
a quantitative theory, however, it is not very successful at lower
densities (which is {\it not} unexpected), and {\it ad hoc}
theoretical refinements (e.g. assuming a lower effective carrier
density or larger effective mass) may be needed for obtaining
quantitative agreement with low density 2D transport data. The
systematic quantitative deviation of the leading-order RPA-Boltzmann
theory from the experimental data on low-density 2D ``metallic''
systems could arise from the large number of effects left out of this
zeroth-order theory which take on significance as the carrier density
is lowered, such as higher-order (beyond screening) interaction
corrections, higher-order impurity scattering effects, possible
localization corrections, and various additional scattering processes
left out of the theory (e.g. intervalley scattering, bulk impurity
scattering, etc.) What is surprising is {\it not} that the
zeroth-order RPA-Boltzmann theory becomes systematically
quantitatively unreliable at lower carrier densities, but the fact
that this minimal leading-order theory provides such an excellent
qualitative description of the observed 2D metallicity (for example,
by explaining the strong variation in $\Delta \rho(T,n)/\rho_0$ in
various systems as arising from the difference in the $q_0$, $t$
values) at all densities and a good quantitative description at
higher densities. This is particularly significant in view of the
early suggestions made in the literature in the context of the 2D
metal-insulator transition physics that the strong 2D metallic
behavior must be arising from some exotic non-Fermi liquid ground
state of the system at low carrier densities. It is now manifestly
clear that the apparent 2D metallic behavior arises from standard
Fermi liquid corrections involved in the interplay of interaction and
disorder with screened effective disorder arising from the temperature
dependent screening of random charged impurities in the system being
the main qualitative source for the strong temperature dependent 2D
resistivity.
This high temperature consequences of the RPA-Boltzmann transport
theory for 2D carrier systems are discussed in our companion
publication \cite{DH11} with the current manuscript focusing entirely
on the low-temperature transport properties.

\section*{ACKNOWLEDGMENTS}

This work is supported by the US-ONR and the NSF-ECS.


\appendix
\section{}

The carrier resistivity $\rho$ in our theory is given by 
\begin{equation}
\rho = \frac{m}{ne^2 \langle \tau \rangle},
\end{equation}
where $m$ is the carrier effective mass, and the energy averaged
transport relaxation time $\langle \tau \rangle$ is given in the
Boltzmann theory by
\begin{equation}
\langle \tau \rangle  = \frac{\int dE
  \tau(E) E \left ( -\frac{\partial f}{\partial E} \right )}{\int dE E \left
    (-\frac{\partial f}{\partial E} \right )},
\end{equation}
where $\tau(E)$ is the energy dependent relaxation time, and $f(E)$ is
the carrier (Fermi) distribution function. At $T=0$, $f(E) \equiv
\theta (E_F-E)$ where $E_F$ is the Fermi energy, and then $\langle
\tau \rangle \equiv \tau(E_F)$ giving the familiar result $\sigma
\equiv \rho^{-1} = ne^2\tau(E_F)/m$. The Fermi energy for a 2D system
is given by 
$E_F = \pi n \hbar^2/(g_v m) \equiv k_B T_F$,
where a spin degeneracy of 2 has been assumed, and $g_v$ is the valley
degeneracy of the 2D system ($g_v=1$ for GaAs; $g_v=2$ for Si (100)
MOSFETs).

We calculate the impurity ensemble averaged relaxation time $\tau(E)$
due to elastic disorder scattering in the Born approximation: 
\begin{eqnarray}
\frac{1}{\tau(E_k)} = \frac{2\pi}{\hbar}\sum_{\alpha,{\bf k}'}
\int^{\infty}_{-\infty}dzN_i^{(\alpha)}(z)
|u^{(\alpha)}({\bf k}-{\bf k}';z)|^2 \nonumber \\
\times (1-\cos \theta_{{\bf k k}'})
\delta(E_k-E_{k'}), 
\label{itau}
\end{eqnarray}
where $E(k) = \hbar^2 k^2/2m$ is the 2D carrier energy for 2D wave
vector {\bf k}; $\theta_{{\bf k k}'}$ is the scattering angle between
carrier scattering wave vectors {\bf k} and ${\bf k}'$; the delta
function $\delta(E_k-E_{k'})$ assures energy conservation for
elastic scattering due to charged impurities where the screened
scattering potential is denoted by $u^{(\alpha)}({\bf q};z)$ 
with ${\bf q} \equiv
{\bf k} - {\bf k}'$ as the 2D scattering wave vector and $z$ is the
quantization or the confinement direction normal to the 2D layer. The
quantity $N_i^{(\alpha)}(z)$ in Eq. (\ref{itau}) denotes the 3D charged
impurity density (with the $z$ dependence reflecting a possible
impurity distribution) of the $\alpha$-th kind with $\alpha$
representing the various possible types of impurities which may be
present in 2D semiconductor structures. For example, $\alpha$ could
denote charged impurities (or interface roughness) located at the
interface (e.g. Si-SiO$_2$ interface for MOSFETs, GaAs-AlGaAs
interface for GaAs heterostructures and quantum wells) or impurities
in the 2D layer itself or remote charged impurities inside the
insulator (which could nevertheless scatter the 2D carriers by virtue
of the long-range nature of Coulomb scattering). It is, for example,
known that in low density Si inversion layers the dominant scattering
arises from the charged impurities located at the Si-SiO$_2$
interface. In general, however, the distribution of scattering centers
in 2D semiconductor systems is {\it not} known, and has to be inferred
from a careful comparison between experimental transport data and
theoretical calculations assuming various kinds of
scatters. Fortunately for our purpose, this lack of precise knowledge
of the charged impurity distribution is not a crucial issue since, on
the qualitative level of our interest, all of them give rise to strong
temperature dependence of the resistivity provided the carrier density
is low enough -- the temperature dependence is the strongest when the
impurities are distributed in the bulk of the 2D layer (bulk
impurities) and goes down as the impurities are located further
(remote impurities) from the layer inside the insulator. In our
calculation presented in this paper, we typically assumes only one
kind of scatters parameterized by a single impurity density to keep
the number of parameters a minimum -- this impurity density
essentially sets the overall scale of resistivity in our results. We
emphasize that we can obtain good qualitative agreement with almost
all the existing 2D MIT experimental data by choosing three different
kinds of charged impurities (i.e. interface, remote, and bulk)
parameterized by a few reasonable parameters, but we do not see much
point in this data fitting-type endeavor.

The screened impurity potential $u^{(\alpha)}({\bf q};z) \equiv
u^{(\alpha)}(q;z)$ is given by: 
\begin{equation}
u^{(\alpha)}(q;z)\equiv [\epsilon(q)]^{-1} V^{(\alpha)}_{\rm{imp}}(q;z)
\end{equation}
where $V^{(\alpha)}_{\rm imp}$ is the bare potential due to a charged
impurity and 
$\epsilon(q)$ is the carrier dielectric function screening the impurity
potential. The bare potential is given by 
\begin{equation}
V^{(\alpha)}_{\rm imp}(q;z) = \frac{2\pi Z^{(\alpha)}e^2}{\bar\kappa q}
F^{(\alpha)}_{\rm imp}(q;z), 
\end{equation}
where $Z^{(\alpha)}$ is the impurity charge strength, $\bar\kappa$ is the
background (static) lattice dielectric constant, and $F_{\rm imp}$ is a
form factor determined by the location of the impurity and the subband
wavefunction $\psi(z)$ defining the 2D confinement. (We do not show
the explicit form of the form-factor $F_{\rm imp}$ here except to note
that it reduces to $F_{\rm imp} = e^{-qd}$ for a strictly 2D layer,
where $|\psi(z)|^2 = \delta(z)$, with $d$ being the separation of the
impurity from the 2D layer -- for $d=0$, i.e. when the impurity is in
the layer, $F_{\rm imp}=1$ in this pure 2D limit as one would expect.)

The finite wave vector dielectric screening function is written in the RPA as 
\begin{equation}
\epsilon(q) = 1- v(q) \Pi(q,T),
\end{equation}
where $v(q)= v_{2D}(q) f(q)$
is the effective bare electron-electron (Coulomb) interaction in the
system with $v_{2D}(q) = 2\pi e^2/(\bar\kappa q)$ being the 2D Fourier
transform of the usual 3D Coulomb potential, $e^2/(\bar\kappa {\bf
  r})$, and
$f(q)$ being the Coulomb form factor arising from the subband
wavefunctions $\psi(z)$:
\begin{eqnarray}
f(q)&=&\frac{1}{2}\int^{\infty}_{-\infty}dz\int^{\infty}_{-\infty}dz'
|\psi(z)|^2|\psi(z')|^2 \nonumber \\
&\times& \left[ \left ( \frac{\kappa_s +
      \kappa_i}{\kappa_s} \right ) e^{-q|z-z'|} + \left (
    \frac{\kappa_s - \kappa_i}{\kappa_s} \right ) e^{-q|z+z'|} \right].
\label{fq}
\end{eqnarray}
The second term in Eq. (\ref{fq}) arises from the image charge effect
due to $\kappa_i \ne \kappa_s$, where $\kappa_i$ and $\kappa_s$ are
the lattice dielectric constants of the insulator and the
semiconductor respectively (with $\bar\kappa = (\kappa_i +
\kappa_s)/2$).
We note that in the strict 2D limit, when $|\psi(z)|^2 = \delta(z)$,
$f(q) = 1$.

The 2D irreducible finite-temperature (and finite wave vector)
polarizability function $\Pi(q;T)$ is given by the noninteracting
polarizability (the irreducible ``bubble'') function within RPA:
\begin{equation}
\Pi(q,T) = \frac{\beta}{4}\int^{\infty}_{0}d\mu'
\frac{\Pi(q;T=0,\mu')}{\cosh^2 \frac{\beta}{2}(\mu-\mu')},
\label{piqt}
\end{equation}
where $\beta \equiv (k_BT)^{-1}$. In Eq. (\ref{piqt}) $\Pi(q;T=0,E_F)$
is the zero-temperature noninteracting static polarizability given by:
\begin{equation}
\Pi(q;T=0,E_F) = N_F \left [ 1- \sqrt{1-\left ( \frac{2k_F}{q}
    \right )^2} \theta(q-2k_F)\right ],
\end{equation}
where $N_F = g_vm/\pi$ is the density of states at Fermi energy, and $k_F
= (2\pi n/g_v)^{1/2}$ is the 2D Fermi wave vector. The chemical
potential $\mu$ in Eq. (\ref{piqt}) at finite temperature $T$ is given
by
\begin{equation}
\mu = \frac{1}{\beta} \ln \left [ -1 + \exp(\beta E_F) \right ].
\end{equation}
We note that the integration in Eq. (\ref{piqt}) is over the dummy
variable $\mu'$ which is unrelated to the real Fermi energy $E_F$ of
the system. For going beyond RPA one would rewrite the noninteracting
polarizability function $\Pi(q,T)$ to a model interacting polarizability
function $\Pi_{\rm int}(q,T)$ which is written as
\begin{equation}
\Pi_{\rm int}(q,T) = \frac{\Pi(q,T)}{1-v(q)G(q,T)\Pi(q,T)},
\end{equation}
where $\Pi$ is the noninteracting polarizability function described
above and $G(q,T)$ is a suitable local field correction which
approximately incorporates correction effects neglected in RPA.

It is now straightforward to see that under the conditions of strict
2D approximation, no remote impurity scattering, only one kind of
impurity scattering (i.e. only one value of $Z_i$ and $N_i$
characterizing the impurity scattering strength), and no local field
corrections the calculated resistivity $\rho(T,n)$ of the system
expressed as the dimensionless quantity $\rho/\rho_0$ where $\rho_0
\equiv \rho(T=0)$ depends only on the variables $q_0 \equiv
q_{TF}/2k_F$ (with the 2D Tomas-Fermi wave vector $q_{TF} =
2g_vme^2/(\bar\kappa \hbar^2)$) and $t \equiv T/T_F$, where
\begin{subequations}
\begin{eqnarray}
q_0& \propto & g_v^{3/2}\frac{m}{\bar \kappa}n^{-1/2}, \\
t  & \propto & T(g_vm)n^{-1}.
\end{eqnarray}
\end{subequations}
For our fully realistic calculations (as well as for the experimental
system), however, this scaling relation, i.e., an exclusive ($q_0,t$)
dependence of resistivity, is violated due to the quasi-2D nature of
the system (i.e. $|\psi(z)|^2 \ne \delta(z)$); the presence of the
insulator (i.e. $\kappa_s \ne \kappa_i$); various types of impurity
distributions in the 2D layer, the interface, and in the insulator;
local field corrections, etc. We note that our calculation assumes one
subband occupancy, i.e. only the ground 2D subband $\psi(z)$ is
considered in our work. At higher temperatures (and lower densities)
other (excited) subbands may get occupied by carriers in which case
one would have to carry out a multisubband transport calculation
including intersubband scattering processes. Such a multisubband
generalization of the Drude-Boltzmann formalism given above is
straightforward, but the actual calculation of $\rho(T)$ becomes
extremely complicate in this situation, and has only been attempted
recently in one special case by us \cite{hwang03} where we considered
intersubband scattering between spin-split subbands in the valance
band of p-GaAs 2D hole systems.

\end{document}